\documentclass[aps,prb,twocolumn,groupedaddress,showpacs,floatfix,eqsecnum,nofootinbib]{revtex4}
\usepackage{graphicx}
\usepackage{latexsym}
\usepackage[tbtags]{amsmath}
\usepackage{amssymb}
\usepackage{bm}
\usepackage{color}
\usepackage{stackrel}

\newcommand{\ddx}[1]{\frac{d #1}{d x}}
\newcommand{\ord}[1]{\bm{\mathit{O}}\left(#1\right)}
\newcommand{\ordSQ}[1]{\bm{\mathit{O}}\left[#1\right]}
\newcommand{\vex}[1]{\bm{\mathrm{#1}}}
\newcommand{\del}{\bm{\nabla}}
\newcommand{\lQ}{l_{Q}}
\newcommand{\lel}{l_{\mathsf{el}}}
\newcommand{\linf}{l_{\infty}}
\newcommand{\lnum}{l_{\mathfrak{r}}^{{\scriptscriptstyle(}\sigma{\scriptscriptstyle)}}}
\newcommand{\lden}{l_{\mathfrak{r}}^{{\scriptscriptstyle(}\kappa{\scriptscriptstyle)}}}
\newcommand{\lnumden}{l_{\mathfrak{r}}^{{\scriptscriptstyle(}\sigma,\kappa{\scriptscriptstyle)}}}

\newcommand{\tauin}{\tau_{\mathsf{in}}}
\newcommand{\tauel}{\tau_{\mathsf{el}}}
\newcommand{\vf}{v_{F}}
\newcommand{\Pres}{\mathcal{P}}
\newcommand{\sig}[1]{\sigma_{#1}}
\newcommand{\Dsig}{D_{\sigma}}
\newcommand{\Sig}[1]{\Sigma^{#1}_{\scriptscriptstyle{I}}}
\newcommand{\BSig}[1]{\Sigma^{#1}_{\scriptscriptstyle{II}}}
\newcommand{\NI}{N_{\scriptscriptstyle{I}}}
\newcommand{\NII}{N_{\scriptscriptstyle{II}}}
\newcommand{\TEP}{\alpha}
\newcommand{\Gth}{G_{\mathsf{th}}}
\newcommand{\RR}{\mathfrak{R}}
\newcommand{\kapinf}{\kappa_{\infty}}
\newcommand{\TEPinf}{\alpha_{\infty}}
\newcommand{\Trsig}{T_{\mathfrak{r}}^{{\scriptscriptstyle(}\sigma{\scriptscriptstyle)}}}
\newcommand{\Trkap}{T_{\mathfrak{r}}^{{\scriptscriptstyle(}\kappa{\scriptscriptstyle)}}}
\newcommand{\Trsigkap}{T_{\mathfrak{r}}^{{\scriptscriptstyle(}\sigma,\kappa{\scriptscriptstyle)}}}

\newcommand{\sigin}{\sigma_{\mathsf{min}}}
\newcommand{\rc}{\mathfrak{r}}
\newcommand{\VwL}{V_{w}^{L}}
\newcommand{\VwO}{V_{w}^{0}}
\newcommand{\Vbi}{V_{\mathsf{bi}}}
\newcommand{\dVwL}{\delta V_{w}^{L}}
\newcommand{\dVwO}{\delta V_{w}^{0}}
\newcommand{\vph}{v_{\mathsf{ph}}}
\newcommand{\tauph}{\tau_{\mathsf{e}\text{-}\mathsf{ph}}}
\newcommand{\lph}{l_{\mathsf{e}\text{-}\mathsf{ph}}}
\begin{document}

\title{
Slow imbalance relaxation and thermoelectric transport in graphene}
\author{Matthew S. Foster}
\email{foster@phys.columbia.edu} 
\author{Igor L. Aleiner}
\affiliation{Department of Physics, Columbia University, New York, NY 10027}
\date{\today}

\begin{abstract}
We compute the electronic component $(\kappa)$ of the thermal conductivity  and the thermoelectric 
power $(\TEP)$ of monolayer graphene, within the hydrodynamic regime, taking into 
account the slow rate of carrier population imbalance relaxation. Interband electron-hole 
generation and recombination processes are inefficient due to the non-decaying nature of 
the relativistic energy spectrum. As a result, a population imbalance of the conduction and 
valence bands is generically induced upon the application of a thermal gradient. We show 
that the thermoelectric response of a graphene monolayer depends upon the ratio of the sample 
length to an intrinsic length scale $\lQ$, set by the imbalance relaxation rate. At the same time, 
we incorporate the crucial influence of the metallic contacts required for the thermopower measurement 
(under open circuit 
boundary conditions), since carrier exchange with the contacts also relaxes the imbalance.
These effects are especially pronounced for clean graphene, where the thermoelectric
transport is limited exclusively by intercarrier collisions. For specimens shorter than $\lQ$,
the population imbalance extends throughout the sample; $\kappa$ and $\TEP$ asymptote toward their 
zero imbalance relaxation limits. In the opposite limit of a graphene slab longer than $\lQ$, at 
non-zero doping $\kappa$ and $\TEP$ approach intrinsic values characteristic of the infinite imbalance 
relaxation limit. Samples of intermediate (long) length in the doped (undoped) case are predicted to 
exhibit an inhomogeneous temperature profile, whilst $\kappa$ and $\TEP$ grow linearly with the 
system size. In all cases except for the shortest devices, we develop a picture of bulk electron and 
hole number currents that flow between thermally conductive leads, where steady-state recombination 
and generation processes relax the accumulating imbalance. Our analysis incorporates, in addition, 
the effects of (weak) quenched disorder.
\end{abstract}

\maketitle

%%%%%%%%%%%%%%%%%%%%%%%%%%%%%%%%%%%%%%%%%%%%%%%%%%%%%%%%%%%%%%%%%%%%%%%%%%%%%%%%%%%%%%%%%%%%%%%%%%%%%%
%%%%%%%%%%%%%%%%%%%%%%%%%%%%%%%%%%%%%%%%%%%%%%%%%%%%%%%%%%%%%%%%%%%%%%%%%%%%%%%%%%%%%%%%%%%%%%%%%%%%%%
%%%%%%%%%%%%%%%%%%%%%%%%%%%%%%%%%%%%%%%%%%%%%%%%%%%%%%%%%%%%%%%%%%%%%%%%%%%%%%%%%%%%%%%%%%%%%%%%%%%%%%

\section{Introduction \label{Sec: Intro}}

Both quenched disorder and interparticle interaction effects influence electric and thermal transport in 
graphene.\cite{CastroNetoReview,NomuraMacDonald,AleinerEfetov,OstrovskyGornyiMirlin1,DasSarma,ClassesAII/A,
CheianovFalkoAltshulerAleiner,DGKP1,Kashuba,SachdevGrapheneQCC,SachdevGrapheneHYDRO} 
At exactly zero doping (the so-called ``Dirac point''), the electron-hole fluid of 
massless Dirac quasiparticles is predicted to exhibit a perfectly finite, non-zero 
dc electrical conductivity $\sigma$ for temperatures $T > 0$, even in the absence of disorder, due entirely to electron-hole 
collisions.\cite{DGKP1,Kashuba,SachdevGrapheneQCC,SachdevGrapheneHYDRO,footnote-a}
Since the carrier plasma is electrically neutral at zero doping, an applied electric field does not couple to 
the center-of-mass momentum of the fluid; instead, electrons and holes are driven in opposite directions, and 
electron-hole collisions limit the developing electric current.
By contrast, a temperature gradient can induce a thermal drift of both electron and hole fluid components in 
the same direction; in the absence of quenched disorder, the resulting energy current grows unimpeded by 
interparticle collisions, which cannot influence the center of mass momentum. 
It has been therefore claimed that the electronic component $\kappa$ of the thermal conductivity of clean, 
undoped graphene is infinite,\cite{SachdevGrapheneHYDRO} being limited only by the amount 
disorder in a dirty sample. We note that the thermoelectric power $\TEP = 0$ at the Dirac point due to 
(approximate) particle-hole symmetry. 

In this paper we demonstrate that the above picture is incomplete: a clean graphene sample at the Dirac point 
will in general exhibit a perfectly finite electronic \emph{thermal conductance} $G_{\mathsf{th}}$, independent 
of the sample length $L$ for $L \gg l_{Q}$, where $l_{Q}$ represents a certain intrinsic length scale 
(set by the bulk properties of the material and the average temperature). In such large specimens, the response 
is predicted to be inhomogeneous, with temperature gradients confined to boundary regions of size $l_{Q}$ adjoining 
the sample edges. In the opposite limit $L \ll l_{Q}$, the temperature falls linearly across the entire sample, and
we find a well-defined, $L$-independent $\kappa$. 
At non-zero doping, the thermoelectric transport properties exhibit several different behavioral regimes.
In the large system 
size limit $L \rightarrow \infty$, both $\kappa$ and $\TEP$ asymptote toward perfectly finite values at non-zero 
$T$ and $\mu$, even in the absence of disorder, 
in accord with the results of Ref.~\onlinecite{SachdevGrapheneHYDRO}.
In the opposite limit of a sufficiently short device $L \ll L_{Q}$, we 
obtain new, completely different results.  The mechanism responsible involves only interparticle collisions, but 
in particular the comparatively slow relaxation of electron-hole population \emph{imbalance}.

\begin{figure}
\includegraphics[width=0.35\textwidth]{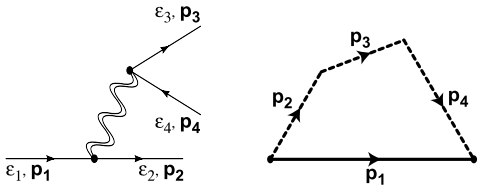}
\caption{
Kinematical constraints for imbalance relaxation. 
The left figure shows the Feynman diagram for the typical two-particle decay process
given by Eq.~(\ref{eDecay}); $\varepsilon_{i}$ and $\vex{p}_{i}$ respectively denote the energy 
and momentum of the $i^{\textrm{th}}$ electron or hole. The right figure depicts momentum conservation 
for this process. The length of the dashed path $L_{f} \equiv |\vex{p}_{2}| + |\vex{p}_{3}| + |\vex{p}_{4}|$ 
traced out by the sum of ``decay product'' momenta is always greater than or equal to the 
length $L_{i} \equiv |\vex{p}_{1}|$ of the ``parent'' particle momentum:  $L_{f} \geq L_{i}$.
If the particle energy spectrum takes the form $\varepsilon(\vex{p}) = |\vex{p}|^{\beta}$, then
the depicted decay process is kinematically forbidden for $\beta < 1$. For $\beta = 1$,
only forward-scattering is allowed.
\label{FigNonDecay}}
\end{figure}

Imbalance refers to a state in which the electron and hole populations deviate from their values in
chemical equilibrium, so that these carriers possess independent chemical potentials $\mu_{e,h}$
and particle densities $n_{e,h}$. 
By contrast, 
equilibrium slaves $\mu_{e} = - \mu_{h} \equiv \mu$ in graphene (a zero bandgap semiconductor), 
so that both $n_{e}$ and $n_{h}$ are completely 
determined by $\mu$ and the temperature $T$. Recombination or generation processes 
relax 
a population imbalance within a time interval $\tau_{Q}$, the imbalance relaxation lifetime.
The lowest order (``Auger'' or two particle collision) relaxation processes typically involve the 
absorption or emission of an electron-hole pair, i.e.\
\begin{subequations}\label{forbidden-processes}
\begin{align}
	e^{-} &\leftrightarrow e^{-} + e^{-} + h^{+},\label{eDecay}
%\\ 
\end{align}
\begin{align}
	h^{+} &\leftrightarrow h^{+} + h^{+} + e^{-}.
\end{align}
\end{subequations}
In graphene, however, these processes are kinematically suppressed
by the conservation of energy $\varepsilon$
and momentum $p$, because the spectrum $\varepsilon(p)$ of the quasiparticles is not
decaying
\[
	\frac{d^2 \varepsilon(p)}{d^2 p} \leq 0.
\]
(\emph{Negative} curvature of the spectrum at $T = 0$ $K$ arises due
to the logarithmic renormalization of the Fermi velocity $v_{F}$ attributed to 
electron-electron interactions.)\cite{AbrikosovBeneslavskii,StauberGuineaVozmediano,Son}
As explicated in Fig.~\ref{FigNonDecay}, the linear spectrum allows only decay products with collinear 
momenta (i.e.\ pure forward scattering), but these processes make a negligible contribution to the 
imbalance relaxation. The sublinear spectrum forbids even this forward scattering decay.
In clean graphene, higher order (e.g.\ three particle collision) imbalance relaxation processes are 
already allowed, while impurity-assisted collisions will contribute in a disordered sample; 
$\tau_{Q}$ is therefore likely finite, although it may significantly exceed other relaxation times 
in the system.

\begin{table*}[t]
\caption{ 
Intrinsic transport parameters for graphene in the hydrodynamic regime, 
and how to extract them from thermoelectric transport (and other) measurements. 
$\kappa(L)$ is the electronic contribution to the thermal conductivity,
and $\alpha(L)$ is the thermopower.
Here and below, $L$ denotes the length of the putative graphene device,
which should be compared to the imbalance relaxation length $\lQ$.
The parameters $\lQ$ and $\sig{eh}$ have been introduced in this work.
Relevant results obtained in this paper can be found in the equations 
listed in the third column. 
\label{TableofPhysObs}
}
\begin{ruledtabular}
\begin{tabular}{l@{\quad}l@{\,}l}
Parameter & Description & Extract from measured quantity\\
\hline 
$\lQ$ 
& Imbalance relaxation length 
& $\bullet$ Temperature $T(x)$ or electrochemical potential $V_{\Phi}(x)$ profile
\\
&
& \phantom{$\bullet$} [Eqs.~(\ref{TmuExp}), (\ref{VPhiDef}), (\ref{PsiSol}), and (\ref{VPhiSol})]
\medskip
\\
$\lel$ 
& Elastic mean free path  
& $\bullet$ Bulk dc conductivity $\sigma$ [Eq.~(\ref{sigmaBulk})], or
\\
& (due to quenched disorder)
& $\bullet$ $\kappa(L) \rightarrow \kapinf$ and $\alpha(L) \rightarrow \TEPinf$ obtained in the limit $L \gg \lQ$
\\
& 
& \phantom{$\bullet$} [Eqs.~(\ref{TCBulk}) and (\ref{TEPBulk})]
\medskip
\\
$\sigin$
& Inelastic carrier-carrier scattering
& $\bullet$ Minimum conductivity at the Dirac point [Eq.~(\ref{sigmaBulk})], or
\\
$\; = \sig{ee}+\sig{hh}-2\sig{eh}$
& contribution to the bulk $\sigma$ 
& $\bullet$ $\kappa(L) \rightarrow \kapinf$ and $\alpha(L) \rightarrow \TEPinf$ obtained in the limit $L \gg \lQ$
\\
&
& \phantom{$\bullet$} [Eqs.~(\ref{TCBulk}) and (\ref{TEPBulk})]
\medskip
\\
$\sig{eh}$
& Off-diagonal (``drag'') component
& $\bullet$ $\kappa(L)$ or $\alpha(L)$ obtained in the limit $L \ll \lQ$
\\
& of the conductivity tensor due to   
& \phantom{$\bullet$} [Eqs.~(\ref{TCshortFinal}) and (\ref{TEPshortFinal})]
\\
& inelastic electron-hole scattering
& 
\end{tabular}
\end{ruledtabular}
\end{table*}

In the limit of zero relaxation, a graphene monolayer probed through thermally conducting, 
electrically insulating contacts would possess electron and hole populations that are strictly conserved. 
In direct analogy with a single component, non-relativistic classical gas,\cite{LLv10} the electron-hole
plasma in clean graphene with vanishing imbalance relaxation would exhibit a finite electronic thermal 
conductivity $\kappa$ for arbitrarily large system sizes. (The \emph{imbalance relaxation length} $l_{Q}$, 
introduced above, diverges as $\tau_{Q} \rightarrow \infty$.) In this regime, interparticle collisions 
facilitate heat conduction without particle number convection. In this paper, we will demonstrate that 
the same behavior obtains for non-zero imbalance relaxation ($1/\lQ > 0$) in the limit of short samples, $L \ll \lQ$. 
By comparison, prior work\cite{SachdevGrapheneHYDRO} effectively assumed infinite relaxation of 
population imbalance ($\lQ \rightarrow 0$). We demonstrate that the results previously obtained in 
Ref.~\onlinecite{SachdevGrapheneHYDRO} for both $\kappa$ and the thermoelectric power $\TEP$ 
at non-zero doping and temperature emerge in the limit of asymptotically large system sizes $L \gg l_{Q}$
for a finite rate of imbalance relaxation, ($\lQ > 0$). We will show that when the ends of a graphene 
slab of length $L \gtrsim \lQ$ are held at disparate temperatures and no electric current is 
permitted to flow, steady state particle convection does nevertheless occur; carrier flux 
is created or destroyed by imbalance relaxation processes near the terminals of the device. 
Thermopower measurements require the junction of the graphene slab with metallic contacts. 
We incorporate into our calculations carrier exchange with non-ideal contacts, which also relax 
the imbalance, 
and we carefully delimit the regimes in which deviations from the infinite 
relaxation limit should be observable in experiments. 
The effects of weak quenched disorder are included in all of our computations. 

In this paper, we restrict our attention to the hydrodynamic (or ``interaction-limited'')
transport regime, where inelastic interparticle collisions dominate over elastic
impurity scattering. Prior work addressing thermoelectric transport in the opposite, 
``disorder-limited'' regime, in which real carrier-carrier scattering processes may be 
neglected, includes that of Refs.~\onlinecite{DGKP1,LofwanderFogelstrom,PeresSantosStauber}.
In the disorder-limited case, $\kappa$ and $\alpha$ are determined
by the energy-dependence of the electrical conductivity, via the ``generalized'' 
Wiedemann-Franz law and Mott relation, respectively.\cite{footnote-b}
The effect of the slow imbalance relaxation upon the dc conductivity in graphene  
under non-equilibrium interband photoexcitation has also been addressed.\cite{Vasko}

What essential new physics emerges through the incorporation of imbalance relaxation
effects into the description of thermoelectric transport, and how can it be extracted from 
experiments?  The entirety of linear transport phenomena in graphene
within the hydrodynamic regime is essentially quantified by four intrinsic parameters.  A finite rate of 
imbalance relaxation means that electrons and holes respond independently to external forces; 
the single ``quantum critical'' conductivity identified previously in 
Refs.~\onlinecite{Kashuba,SachdevGrapheneQCC,SachdevGrapheneHYDRO} generalizes
to a 2x2 tensor of coefficients, with diagonal elements $\sig{ee}$, $\sig{hh}$ and off-diagonal
elements $\sig{eh} = \sig{he}$; all are mediated entirely by inelastic interparticle collisions.
The description is similar to that of Coulomb drag:\cite{Rojo}
the diagonal element $\sig{ee}$ ($\sig{hh}$) characterizes the response of the conduction band 
electrons (valence band holes) to a (gedanken) electric field that couples only to that carrier
type, whereas $\sig{eh}$ characterizes the ``drag'' exerted by one carrier species upon the other
(due to electron-hole collisions) under the application of such a field.
$\sig{ee}$ and $\sig{hh}$ are related by particle-hole symmetry, 
leaving two independent parameters which we can take as $\sig{eh}$ and 
$\sigin \equiv (\sig{ee} + \sig{hh} - 2 \sig{eh})$; 
the latter combination is equal to the bulk dc electrical conductivity $\sigma$ 
at the Dirac point in the hydrodynamic regime.\cite{footnote-a}
The imbalance relaxation length $\lQ$ constitutes the third intrinsic graphene parameter, 
while the fourth is provided by the elastic mean free path $\lel$ for a disordered sample. 

Of these, $\lel$ and $\sigin$ can both be obtained from either the bulk dc conductivity 
$\sigma$ (measured at variable doping), or from the combined measurement of the ``infinite imbalance 
relaxation'' thermoelectric transport coefficients $\kapinf$ and $\TEPinf$. 
Here, $\kapinf$ and $\TEPinf$ are the (electronic) thermal conductivity and thermopower, respectively, 
as measured in a graphene device with $L \gg \lQ$. 
By contrast, the off-diagonal parameter 
$\sig{eh}$ can be independently determined through measurement of either $\kappa$ or $\alpha$, for a 
sufficiently short device satisfying $L \ll \lQ$. (A precise discussion of the limiting and 
crossover behaviors of $\kappa$ and $\alpha$, incorporating the complicating effects of the external 
contacts, is presented in Sec.~\ref{Sec: TCandTEP} of this paper). Finally, the imbalance relaxation 
length $\lQ$ can be ascertained via the measurement of either $\kappa$ or $\alpha$ in the crossover 
regime $L \sim \lQ$, or through spatial 
resolution of the inhomogeneous temperature or electrochemical 
potential profiles across a device with $L \gtrsim \lQ$. The four intrinsic graphene transport parameters
discussed above and suggestions for their experimental determination are listed in 
Table~\ref{TableofPhysObs}.

The rest of this paper is organized as follows.

In Sec.~\ref{Sec: Hydro}, we formulate a hydrodynamic description of carrier transport that admits
carrier population imbalance and relaxation. In order to illustrate ideas, we apply this formalism
to a putative experimental device. The hydrodynamic approach allows us to obtain the inhomogeneous 
temperature, chemical potential, and number current profiles across the device; some of these are 
sketched in Fig.~\ref{TempCurrentDensity} for a system at zero doping.
In Sec.~\ref{Sec: TCDiracPoint}, we derive the thermal conductance $G_{\mathsf{th}}$ at the Dirac point,
and we discuss the short and long device asymptotics of heat transport.   
General results for $\kappa$ and $\TEP$ at arbitrary doping are obtained and discussed in the penultimate
Sec.~\ref{Sec: TCandTEP}. 

We neglect phonon effects throughout the body of this work, but these are considered 
in the Conclusion Sec.~\ref{Sec: Conclusion}. We focus in particular upon the influence of electron-phonon 
interactions on the \emph{electronic} transport coefficients derived in this paper. We demonstrate that 
phonons may be neglected at low temperatures or for graphene samples of mesoscopic size.

%%%%%%%%%%%%%%%%%%%%%%%%%%%%%%%%%%%%%%%%%%%%%%%%%%%%%%%%%%%%%%%%%%%%%%%%%%%%%%%%%%%%%%%%%%%%%%%%%%%%%%
%%%%%%%%%%%%%%%%%%%%%%%%%%%%%%%%%%%%%%%%%%%%%%%%%%%%%%%%%%%%%%%%%%%%%%%%%%%%%%%%%%%%%%%%%%%%%%%%%%%%%%
%%%%%%%%%%%%%%%%%%%%%%%%%%%%%%%%%%%%%%%%%%%%%%%%%%%%%%%%%%%%%%%%%%%%%%%%%%%%%%%%%%%%%%%%%%%%%%%%%%%%%%

\section{Hydrodynamic formulation and solution \label{Sec: Hydro}}

\subsection{Two fluid hydrodynamics}

We take as our starting point the hydrodynamic equations of motion for carriers in graphene, 
expressed in relativistically covariant notation: 
\begin{gather}
	\partial_{i} J^{i}_{e}
	= 
	- \partial_{i} J^{i}_{h}
	= 
	- e I,
	\label{ParticleCons}
	\\
	\partial_{j} \Theta^{i j} - \frac{1}{v_{F}}F^{i j} (J_{e\,j} + J_{h\,j})
	=
	b_{\mathsf{el}}^{i},
	\label{EnergyMomCons}
\end{gather}
where $J_{e}^{i}$ and $J_{h}^{i}$ denote the electron and hole electric 3-current densities, 
respectively, $\Theta^{i j}$ is the traceless energy-momentum tensor for the (classically) 
scale-invariant two-component plasma, and $F^{i j}$ is the Faraday tensor, incorporating 
both external and self-consistent fields. 
The Fermi velocity in graphene is denoted by $\vf$, while $e = -|e|$ is the electron charge.
Summation over repeated ``spacetime'' indices is implied in Eqs.~(\ref{ParticleCons}) and 
(\ref{EnergyMomCons}), where $x^{i} \in \{x^{0},x^{1},x^{2}\} = \{\vf t, x, y\}$, while 
$\partial_{i} \equiv \partial/\partial x^{i}$. The quantity $I$ in Eq.~(\ref{ParticleCons}) 
is the \emph{imbalance relaxation flow}, which 
is proportional to the rate $1/\tau_{Q}$ of carrier recombination or generation between the 
electron and hole bands. The frictional force density $b_{\mathsf{el}}^{i}$ in Eq.~(\ref{EnergyMomCons}) 
manifests the effects of elastic scattering of carriers by quenched disorder. In a microscopic 
quantum kinetic equation treatment, $I$ and $b_{\mathsf{el}}^{i}$ obtain as certain momentum 
averages of the inelastic and elastic collision integrals, respectively.
In the limit $I \rightarrow 0$, electrons and holes are separately conserved.

By adopting the hydrodynamic approach, we posit fast equilibration of the electron-hole
plasma due to inelastic collisions; in particular, we assume $\tauin \lesssim \tauel$
(the ``interaction-limited'' transport regime, see Ref.~\onlinecite{DGKP1}),
with $\tauin$ the inelastic lifetime due to interparticle scattering, and $\tauel$ the 
elastic lifetime due to quenched disorder. We express $J_{e,h}^{i}$ and $\Theta^{i j}$ in terms
of local thermodynamic variables and a hydrodynamic 3-velocity 
$U^{i}\equiv\gamma(v_{F},\vex{u})$, with $\vex{u}$ the ordinary fluid velocity and 
$\gamma^2 = (1 - \vex{u}^2/v_{F}^2)^{-1}$.
Incorporating dissipative deviations from local equilibrium, we write\cite{LLv6}
\begin{align}
	J^{i}_{e} 
	&\equiv 
	e \left(n_{e} U^{i} + \nu_{e}^{i}\right),
	\label{Jiedecomp}
	\\
	J^{i}_{h} 
	&\equiv 
	-e \left(n_{h} U^{i} + \nu_{h}^{i}\right),
	\label{Jihdecomp}
	\\
	\Theta^{i j}
	&\equiv
	3 \mathcal{P}
	\left(
	\frac{1}{v_{F}^2}U^{i} U^{j} - \frac{1}{3}g^{i j}
	\right)
	+ \theta^{i j},
	\label{Thetaijdecomp}
\end{align}
where $\nu_{e,h}^{i}$ and $\theta^{i j}$ define dissipative fluctuations of the electron and hole
number currents and stress tensor, respectively. In Eqs.~(\ref{Jiedecomp}) and (\ref{Jihdecomp}), 
$n_{e}$ and $n_{h}$ represent proper (rest frame) electron and hole densities. $\mathcal{P}$ 
denotes the total pressure in Eq.~(\ref{Thetaijdecomp}), where we have used the thermodynamic 
relation $3 \mathcal{P} = \mathfrak{h}$, with $\mathfrak{h}$ 
the enthalpy density; this is a 
consequence of relativistic scale invariance. In this same equation, $g^{i j}$ denotes the 
Minkowski metric tensor.\cite{footnote-c}

In what follows, we will neglect for compactness of presentation the non-equilibrium 
component of the stress tensor $\theta^{i j}$, which describes viscous effects in 
non-uniform fluid flow. 
For the thermoelectric transport problem studied here,
viscous drag along the sample edges is irrelevant under the condition 
$W^2 \gg \eta \vf^2 \tauel/3 \Pres$, where $W$ is the sample width transverse to parallel
electric and heat current flows, and $\eta$ is the first (dynamic) viscosity coefficient.\cite{LLv6}
(The second viscosity $\zeta$ vanishes for a massless relativistic gas.)\cite{LLv10} 
By inserting the decomposition in Eqs.~(\ref{Jiedecomp})--(\ref{Thetaijdecomp}) into 
Eqs.~(\ref{ParticleCons}) and (\ref{EnergyMomCons}), and taking the non-relativistic
limit $\vex{u}^2 \ll v_{F}^2$, we derive the following entropy, momentum, and particle 
number balance equations:
\begin{align}\label{EntropyBalance}
	&\begin{aligned}[b]
	T
	\bigg[
	\partial_{t} s
	&+ \del\cdot\left(
	s \vex{u} - \frac{\mu_{e}}{T}\vex{\nu}_{e} - \frac{\mu_{h}}{T}\vex{\nu}_{h}
	\right)
	\bigg]
	\\
	&
	\begin{aligned}[b]
	=&
	\left[
	e \vex{E} - T \del \left(\frac{\mu_{e}}{T}\right)
	\right]
	\cdot\vex{\nu}_{e}
	+
	(\mu_{e}+\mu_{h})I
	\\
	&
	+
	\left[
	- e \vex{E} - T \del \left(\frac{\mu_{h}}{T}\right)
	\right]
	\cdot\vex{\nu}_{h}
	-
	\vex{u}\cdot\vex{b}_{\mathsf{el}},
	\end{aligned}
	\end{aligned}
	\\
	\label{NavierStokes}
	&\frac{3 \mathcal{P}}{v_{F}^2}
	\frac{d \vex{u}}{d t}
	=
	\rho \vex{E}
	- n_{e} \del \mu_{e}
	- n_{h} \del \mu_{h}
	- s \del T
	+\vex{b}_{\mathsf{el}}
	,
	\\
	\label{NumberImbalance}
	&\partial_{t} n
	+ \del\cdot\vex{J}_{n}
	= - 2 I,
\end{align}
\begin{align}
	\label{ChargeContinuity}
	&\partial_{t} \rho
	+ \del\cdot\vex{J}
	=0.
\end{align}
Here, $s$ is the entropy density, $\mu_{e,h}$ are the electron and hole chemical potentials,
and $\vex{E}$ is the electric field. $\vex{J}_{n}$ and $\vex{J}$ respectively denote the 
total carrier number and electric current densities in Eqs.~(\ref{NumberImbalance}) and 
(\ref{ChargeContinuity}); these are defined as 
\begin{align}  
	\vex{J}_{n} &\equiv n \vex{u} + \vex{\nu}_{e} + \vex{\nu}_{h},
	\label{JnDef}
	\\
	\vex{J} &= \vex{J}_{e} + \vex{J}_{h}
	\nonumber\\
	&=
	\rho \vex{u} + e (\vex{\nu}_{e}-\vex{\nu}_{h}).
	\label{JDef}
\end{align}
In Eqs.~(\ref{NumberImbalance})--(\ref{JDef}),
\begin{align}\label{nrhoDef}
	n \equiv n_{e} + n_{h},\quad
	\rho \equiv e(n_{e} - n_{h})	
\end{align}
represent the net carrier number and electric charge densities, respectively.
Let us also define the \emph{imbalance} $\mu_{I}$ and \emph{relative}
$\mu$ chemical potentials, via
\begin{equation}\label{muDefs}
	\mu_{I} \equiv \frac{\mu_{e} + \mu_{h}}{2},\quad
	\mu \equiv \frac{\mu_{e} - \mu_{h}}{2}.
\end{equation}
On the left-hand side of Eq.~(\ref{NavierStokes}), 
\begin{equation}
	\frac{d}{d t} \equiv \partial_{t} + \vex{u}\cdot\del
\end{equation}
is the material derivative.

Within linear response,
Eq.~(\ref{EntropyBalance}) implies that the thermodynamic ``forces'' 
$\{
	e \vex{E} - T \del (\mu_{e}/{T}),
	- e \vex{E} - T \del (\mu_{h}/{T}),
	\vex{u},
	2 \mu_{I}
\}$
determine the conjugate ``fluxes'' 
$\{\vex{\nu}_{e,h},\vex{b}_{\mathsf{el}},I\}$ via the matrix equation
\begin{equation}\label{LinearResponse1}
	\begin{bmatrix}
	e^2 \vex{\nu}_{e}\\
	e^2 \vex{\nu}_{h}\\
	\vex{b}_{\mathsf{el}}\\
	I
	\end{bmatrix}
	=
	\hat{\mathrm{M}}
	\begin{bmatrix}
	e \vex{E} - T \del \left(\frac{\mu + \mu_{I}}{T}\right)\\
	- e \vex{E} + T \del \left(\frac{\mu - \mu_{I}}{T}\right)\\
	\vex{u}\\
	2 \mu_{I}
	\end{bmatrix}.
\end{equation} 
Onsager reciprocity\cite{deGroot} of the entropy balance Eq.~(\ref{EntropyBalance}) 
dictates that the kinetic coefficient matrix $\hat{\mathrm{M}}$ is symmetric, 
\[
	\hat{\mathrm{M}}^{\mathsf{T}} = \hat{\mathrm{M}}, 
\]
where the superscript ``$\mathsf{T}$'' denotes the matrix transpose operation. Assuming 
vanishing or weak quenched disorder and slow imbalance relaxation, $\hat{\mathrm{M}}$
can be written as
\begin{equation}\label{LinearResponse2}
	\hat{\mathrm{M}}
	=
	\begin{bmatrix}
	\sigma_{ee} & \sigma_{eh} & 0 & 0\\
	\sigma_{eh} & \sigma_{hh} & 0 & 0\\
	0 & 0 & -\frac{3 \mathcal{P}}{v_{F}^2 \tauel} & 0\\
	0 & 0 & 0 & \frac{n \lambda_{Q}}{2 \hbar}
	\end{bmatrix}.
\end{equation}
The electric conductivities $\sigma_{\mathsf{ab}}$, $\mathsf{ab} \in \{ee,hh,eh\}$, arise solely due
to interparticle collisions, and can be computed in principle within a microscopic 
quantum kinetic equation (QKE) formulation.\cite{footnote-a} The elastic lifetime 
$\tauel$ determines the frictional force density $\vex{b}_{\mathsf{el}} \propto - \vex{u}$; 
Eqs.~(\ref{LinearResponse1}) and (\ref{LinearResponse2}) provide an implicit definition for 
$\tauel$. Finally, $\lambda_{Q}$ is a dimensionless parameter that characterizes the 
efficacy of generation and recombination processes. 

Both $\{\sigma_{\mathsf{ab}}\}$ and $\lambda_{Q}$ are functions of the dimensionless ratios 
$\mu/k_{B} T$ and $\mu_{I}/k_{B}T$. Particle-hole symmetry requires that the diagonal 
conductivity elements satisfy the condition
\begin{equation}\label{PHSymsigma}
	\sig{ee}\left(\frac{\mu}{k_{B} T},\frac{\mu_{I}}{k_{B} T}\right)
	=
	\sig{hh}\left(-\frac{\mu}{k_{B} T},\frac{\mu_{I}}{k_{B} T}\right).
\end{equation}
Eqs.~(\ref{LinearResponse1}) and (\ref{LinearResponse2}) therefore assert 
that the interparticle collisions give rise to \emph{two} independent kinetic 
coefficients $\sig{ee}$ and $\sig{eh}$ that carry the units of electrical conductance, 
in addition to the dimensionless imbalance relaxation parameter $\lambda_{Q}$.
In this paper we focus upon transport in the non-degenerate regime, $k_{B} T \gg |\mu|, |\mu_{I}|$;
within the accuracy of the linear response approximation, 
$\{\sigma_{\mathsf{ab}}\}$ and $\lambda_{Q}$ can then be regarded as fixed constants, 
typically evaluated at the Dirac point ($\mu = \mu_{I} = 0$). Under these conditions,
Eq.~(\ref{PHSymsigma}) implies that $\sig{ee} = \sig{hh}$.

It is worth pointing out that the friction density $\vex{b}_{\mathsf{el}}$ and 
imbalance relaxation flow $I$ appear already at the level of the ``ideal'' hydrodynamics: 
to compute (in principle) the parameters $1/\tauel$ and $\lambda_{Q}$ in 
Eq.~(\ref{LinearResponse2}), it is sufficient to solve the associated QKE at \emph{zeroth} 
order in the inelastic relaxation time $\tauin$, with electron and hole distribution 
functions locally constrained to take the Fermi-Dirac form.\cite{LLv6,LLv10,Uhlenbeck} 
The ideal hydrodynamics is captured by Eqs.~(\ref{EntropyBalance})--(\ref{JDef}) with 
$\vex{\nu}_{e} = \vex{\nu}_{h} = 0$. 

To zeroth order in $\tauin$, $\vex{b}_{\mathsf{el}}$ is nonvanishing for a convective 
particle flow in the rest frame of the disorder ($\vex{u} \neq 0$), while nonzero $I$ 
arises whenever a population imbalance occurs ($\mu_{I} \neq 0$).
The explanation for this is as follows: 
the inelastic collision integral governing interparticle scattering for a clean graphene 
system with vanishing imbalance relaxation would possess zero modes associated to homogeneous 
fluid convection (momentum conservation) and global shifts of the total carrier number density. 
These zero modes are made ``massive'' through the introduction of disorder and the inclusion of \emph{some} 
mechanism for imbalance relaxation (such as three particle collisions). 
For graphene in the hydrodynamic regime, the ``masses'' (scattering rates) ($1/\tauel$) and
($1/\tau_{Q}$) respectively associated to \emph{weak} disorder and \emph{inefficient} imbalance 
relaxation are much smaller than inelastic scattering rate ($1/\tauin$) for electron and hole 
number-conserving collisions. Even in the limit of arbitrarily efficient equilibration due to 
frequent and strongly inelastic such collisions ($\tauin \rightarrow 0$), nonvanishing $\vex{u}$ 
and $\mu_{I}$ induce dissipation by coupling to these weakly massive modes. 
The dominant effect of these forces is the generation of $\vex{b}_{\mathsf{el}}$ and $I$ 
in the ideal hydrodynamic description. 

By contrast, a small but non-zero 
$\tauin$ allows out-of-equilibrium deformations of the electron and hole distribution function shapes. 
The $\{\sigma_{\mathsf{ab}}\}$ acquire nonzero values in the first order of the QKE expansion in $\tauin$. 
This separation of zeroth and first order responses justifies the assumed block-diagonal form of 
the kinetic coefficient matrix in Eq.~(\ref{LinearResponse2}).

\subsection{``Bulk'' kinetic coefficients\label{BulkKinetics}}

In the limit of \emph{infinite} imbalance relaxation, $\lambda_{Q} \rightarrow \infty$ in
Eq.~(\ref{LinearResponse2}), we must slave $\mu_{e} = - \mu_{h} = \mu$ and $\mu_{I} = 0$ 
in Eqs.~(\ref{EntropyBalance}), (\ref{NavierStokes}), and (\ref{LinearResponse1}). 
Assuming steady-state conditions, Eqs.~(\ref{NavierStokes})--(\ref{ChargeContinuity}), 
(\ref{LinearResponse1}), and (\ref{LinearResponse2}) allow the computation of the electric 
$\vex{J}$ and heat $\vex{J}_{q}$ current densities in the presence of electrochemical potential and 
temperature gradients. The heat current implied by the left-hand side of the entropy balance 
Eq.~(\ref{EntropyBalance}) is
\begin{align}  
	\vex{J}_{q} 
	&\equiv 
	T s \vex{u} - \mu_{e} \vex{\nu}_{e} - \mu_{h} \vex{\nu}_{h}
	\nonumber\\
	& = 
	3\mathcal{P} \vex{u} - \mu_{I} \vex{J}_{n} - \frac{\mu}{e} \vex{J},
	\label{JqDef}
\end{align}
where on the second line we have used Eqs.~(\ref{JnDef}) and (\ref{JDef}).
The linear response takes the usual form\cite{AshcroftMermin}
\begin{align}
	\vex{J} 
	&=
	\sigma \vex{\varepsilon}
	+ \sigma \TEPinf \left(-\del T\right),
	\label{JBulk}\\
	\frac{1}{T} \vex{J}_{q} 
	&=
	\sigma \TEPinf \vex{\varepsilon}
	+ \left(\frac{\kapinf}{T} + \sigma \TEPinf^2 \right)
	\left(- \del T \right),
	\label{JqBulk}
\end{align}
where
\[
	\vex{\varepsilon} \equiv \vex{E} - \frac{1}{e}\del\mu
\]
is the electrochemical potential gradient.
The thermoelectric response in Eqs.~(\ref{JBulk}) and (\ref{JqBulk})
is characterized by the bulk dc electrical conductivity $\sigma$,
and the infinite imbalance relaxation limits of the thermopower $\TEPinf$ and
thermal conductivity $\kapinf$. In terms of the intrinsic kinetic parameters 
defined via 
Eq.~(\ref{LinearResponse2}), one finds that
\begin{align}
	\sigma 
	&=
	\sigin
	+ \frac{\vf \lel \rho^2}{3 \Pres},
	\label{sigmaBulk}\\
	\kapinf
	&=
	\frac{3 \Pres \vf \lel \sigin}
	{T \sigma},
	\label{TCBulk}\\
	\TEPinf
	&=
	\frac{\vf \lel \rho}{T \sigma} - \frac{\mu}{e T},
	\label{TEPBulk}
\end{align}
where the minimum conductivity at the Dirac point is given by
\begin{equation}\label{siginDef}
	\sigin \equiv \sig{ee} + \sig{hh} - 2 \sig{eh}.
\end{equation}
In these equations, $l_{\mathsf{el}} \equiv v_{F} \tauel$ is the elastic mean free 
path due to quenched impurity scattering. 
In the clean limit, the thermopower in Eq.~(\ref{TEPBulk}) simplifies to $\TEPinf = s/\rho$, 
which may be interpreted as the ``transport entropy'' per charge.\cite{Domenicali}
The results of Eqs.~(\ref{sigmaBulk})--(\ref{TEPBulk})
were originally obtained in Ref.~\onlinecite{SachdevGrapheneHYDRO}. 

All three thermoelectric coefficients in Eqs.~(\ref{sigmaBulk})--(\ref{TEPBulk})
depend only upon the combination $(\sig{ee} + \sig{hh} - 2 \sig{eh})$, for arbitrary doping.
We will demonstrate below that these same formulae apply in the presence of finite
imbalance relaxation, in the limit of a sufficiently 
long graphene sample. In the opposite limit of a short graphene device, we obtain new, completely different
results for both $\alpha$ and $\kappa$, as detailed in Sec.~\ref{Sec: TCandTEP}. 
Combined with particle-hole symmetry [Eq.~(\ref{PHSymsigma})],
measurement of long and short graphene devices should allow an experimental determination
of all three parameters $\sig{ee}$, $\sig{hh}$, and $\sig{eh}$.

\subsection{Experimental geometry}

\begin{figure}
\includegraphics[width=0.3\textwidth]{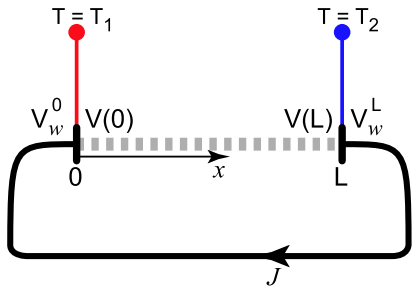}
\caption{
Schematic setup for the two terminal thermopower and thermal conductivity
measurements. The graphene slab of length $L$ is denoted by the dashed gray
line. The contacts at $x = 0,L$ are respectively held at temperatures 
$T = T_{1}, T_{2}$ via external thermostats. The two contacts are electrically
bridged by a highly resistive wire, represented in the figure by the thick black 
interconnect. The development of a thermoelectric response $V(L) - V(0) \neq 0$ in 
the graphene induces a voltage drop $\Delta V_{w} \equiv V_{w}^{L} - V_{w}^{0}$ in the wire, 
which drives the current $J$. Measurement of $J$ (via galvanometer) and knowledge of
the wire resistance allows computation of the graphene thermopower, which is simply the
ratio of $\Delta V_{w}$ to the temperature drop $\Delta T = T_{1} - T_{2}$, in the limit
of infinite wire resistance.  
\label{FigSetup}}
\end{figure}

We consider an experiment in which a rectangular strip of
graphene is terminated with metallic contacts at opposite ends of the strip; we take the 
strip to lie along the x-axis between $x = 0$ and $x = L$. In a thermal conductivity measurement, 
the leads are held at different temperatures, $T(0) \equiv T_{1}$ and $T(L) \equiv T_{2}$, and 
the heat current $\vex{J}_{q}$ [Eq.~(\ref{JqDef})] is measured. 
For convenience, let us introduce
\begin{equation}\label{TempAvgandDiff}
	\bar{T} \equiv \frac{T_{1}+T_{2}}{2},\quad 
	\Delta T \equiv T_{1} - T_{2}.
\end{equation}
To determine the thermopower, we analyze a gedanken measurement in which
the metallic contacts are interconnected by a highly resistive ``wire'';
we sketch a schematic setup in Fig.~\ref{FigSetup}. The wire is taken to 
be composed of a disordered metal that is approximately particle-hole symmetric, 
and thus manifests no thermoelectric voltage of its own. Let $V(x)$ denote the electric 
potential profile within the graphene. The application of the temperature gradient $\Delta T \neq 0$
induces a drop $V(L) - V(0)$ across the graphene slab via its thermoelectric effect. 
Through local electrochemical quasiequilibration at the contacts, the graphene potential 
drop translates into a voltage difference $\Delta V_{w} \equiv \VwL - \VwO$ across the 
ends of the wire. Here, $\VwL$ and $\VwO$ denote the electric potentials near the 
contacts situated at $x = L$ and $x = 0$, respectively, \emph{just inside of the metal 
wire} (outside of the graphene). $\Delta V_{w}$ drives an electric current through 
the wire linking the contacts. We assume local electroneutrality throughout the wire; 
therefore, the diffusion component of the electric current vanishes outside of the graphene. 
In the limit of arbitrarily large wire resistance, the thermopower is then simply the ratio
\begin{equation}\label{TEPDef}
	\TEP \equiv \frac{\Delta V_{w}}{\Delta T}.
\end{equation}
Knowledge of the wire resistance and a galvanic measurement of the current thus allow
experimental determination of $\TEP$.

We specialize to the quasi-1D strip geometry discussed above, and assume a steady-state 
linear response to the application of unequal temperatures at the contacts. A small $\Delta T$
induces proportional deviations from zero of $u$, $\mu_{I}$, $J_{n}$, and $V(x)$; we define
the graphene electric potential such that $V(x) \rightarrow 0$ in the limit $\Delta T \rightarrow 0$.
The temperature and relative chemical potential are expanded about their average values:
\begin{equation}\label{TmuExp}
	T(x) = \bar{T} + \Psi(x),\quad 
	\mu(x) = \bar{\mu} + \Phi(x),
\end{equation}
where $\Psi$ and $\Phi$ are assumed proportional to $\Delta T$.\\

\subsection{Boundary conditions}

\begin{figure}
\includegraphics[width=0.45\textwidth]{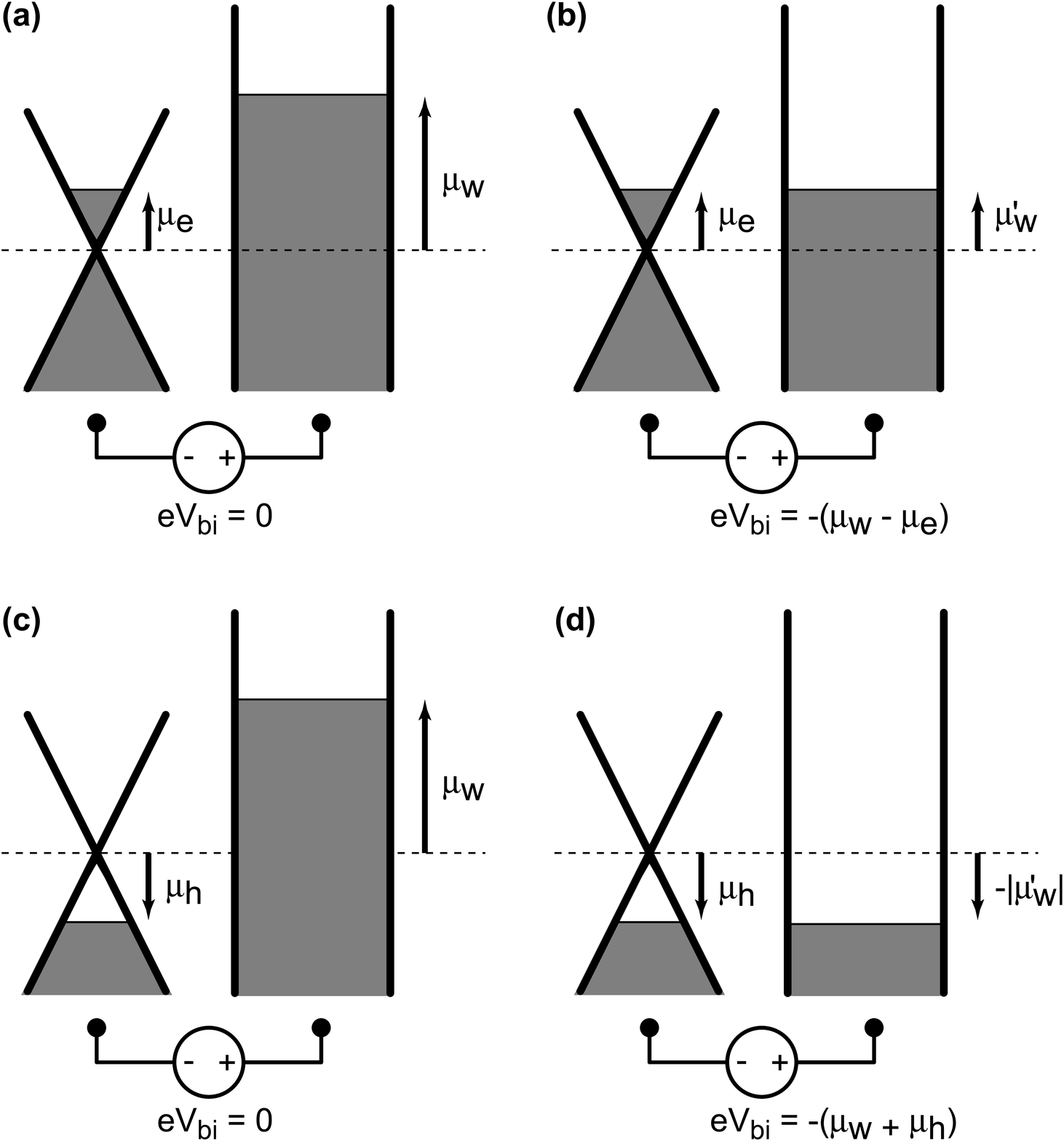}
\caption{
This figure depicts contact (``built-in'') potentials $\Vbi$ that arise 
through equilibration of graphene and the wire interconnect (voltmeter)
depicted in Fig.~\ref{FigSetup}. Subfigures {\bf(a)} and {\bf(b)} depict
the electrochemical equilibration of electron-doped graphene and the metallic interconnect,
while {\bf(c)} and {\bf(d)} show the same for the hole-doped case. All chemical
potentials are measured from the position of the Dirac point; in this figure,
we have assumed that $\mu_{w} > 0$, i.e.\ the work function of the metal is less than
the electron affinity of the graphene. The notion of a ``negative hole chemical potential'' 
$- \mu_{w}$ for the carriers in the wire is easily understood via subfigure {\bf(c)}: 
before electrochemical equilibration, the chemical potential difference $\mu_{h} - (- \mu_{w})$ 
drives holes in the graphene to ``float up'' toward the hole Fermi sea in the metal. 
Given the assumption of local electroneutrality throughout both the graphene and metal, 
the tunneling holes must recombine with electrons on the surface of the metal, leading 
to the accumulation of a dipole charge layer at the boundary. As a result, a static 
built-in voltage $\Vbi = - \frac{1}{e}(\mu_{w} + \mu_{h})$ develops between the 
wire and graphene which precisely offsets the intrinsic chemical potential difference. 
Of course, this built-in voltage is \emph{not} directly measurable with a voltmeter.
\label{FigChemPot}}
\end{figure}

We impose the following boundary conditions\cite{Konin} relating the electron and hole electric current
densities $J_{e,h}$ to the corresponding electrochemical potential drops across the contacts of the device 
shown in Fig.~\ref{FigSetup}:
\begin{subequations}\label{CurrentContBC}
\begin{align}
	J_{e}(L) 
	&= \frac{1}{2 \rc} \left\{V(L) - \VwL - \frac{1}{e}\left[\mu_{w} - \mu_{e}(L)\right]\right\}, 
	\label{CurrentContBC1}\\
	J_{h}(L)
	&= \frac{1}{2 \rc} \left\{V(L) - \VwL + \frac{1}{e}\left[-\mu_{w} - \mu_{h}(L)\right]\right\},
	\label{CurrentContBC2}\\
	J_{e}(0)
	&= \frac{1}{2 \rc} \left\{\VwO - V(0) - \frac{1}{e}\left[\mu_{e}(0) - \mu_{w}\right]\right\},
	\label{CurrentContBC3}\\
	J_{h}(0)
	&= \frac{1}{2 \rc} \left\{\VwO - V(0) + \frac{1}{e}\left[\mu_{h}(0) + \mu_{w}\right]\right\}.
	\label{CurrentContBC4}
\end{align}
\end{subequations}
Here, $\rc$ denotes the contact surface resistivity, i.e.\ $R_{c} = \rc/W$,
where $R_{c}$ is the electrical contact resistance and $W$ is the sample width 
transverse to the temperature gradient. In Eqs.~(\ref{CurrentContBC1})--(\ref{CurrentContBC4}), 
$\mu_{w}$ denotes the single chemical potential level characterizing the particle-hole 
symmetric metal wire. Given the assumption of electroneutrality, $\mu_{w}$ remains fixed 
at its equilibrium value even in the presence of $\Delta T \neq 0$. 

In equilibrium, $T_{1} = T_{2} = \bar{T}$; everywhere within the graphene slab,
we have the conditions
\begin{gather}
	\mu_{e}(x) = - \mu_{h}(x) = \bar{\mu},\quad 
	V(x) = 0,
	\nonumber\\
	J_{e}(x) = J_{h}(x) = 0.
	\nonumber
\end{gather}
Generically, the electron chemical potentials in the graphene slab ($\mu_{e} = \bar{\mu}$) 
and metal wire ($\mu_{w}$) will differ; equivalently, 
$\mu_{h} - (-\mu_{w}) = - \bar{\mu} + \mu_{w} \neq 0$. The interpretation of $\mu_{w}$ and $-\mu_{w}$
as the ``electron'' and ``hole'' chemical potentials in the metal is elaborated in Fig.~\ref{FigChemPot}. 
Assuming strong bulk screening in both materials, a dipole layer develops at the contacts, leading to the
contact potential (see Fig.~\ref{FigChemPot})
\[
	\VwL = \VwO = \frac{1}{e}(\bar{\mu} - \mu_{w}).
\] 
In the nonequilibrium case, we therefore write
\begin{align}\label{dVwDefs}
	\VwL = \frac{1}{e}(\bar{\mu} - \mu_{w}) + \dVwL,\quad
	\VwO = \frac{1}{e}(\bar{\mu} - \mu_{w}) + \dVwO.
\end{align}

We now take the ideal limit of an infinitely resistive wire interconnect (Fig.~\ref{FigSetup}),
and thus we require that the electric current vanish everywhere:
\[
	J = 0.
\]
Using Eqs.~(\ref{JnDef}), (\ref{JDef}), (\ref{muDefs}),  and (\ref{dVwDefs}),
the boundary conditions in Eq.~(\ref{CurrentContBC}) may then be recast as
\begin{subequations}\label{BCFinal}
\begin{gather}
	\mu_{I}(L)
	=
	e^2 \, \rc \, J_{n}(L),\quad
	- 
	\mu_{I}(0)
	=
	e^2 \, \rc \, J_{n}(0),
	\label{BCmuI}\\
	\dVwL
	=
	V_{\Phi}(L),\quad
	\dVwO
	=
	V_{\Phi}(0)
	\label{BCdVwL},
\end{gather}
\end{subequations}
where we have introduced the electrochemical potential fluctuation
\begin{equation}\label{VPhiDef}
	V_{\Phi}(x) \equiv V(x) + \frac{1}{e} \Phi(x).
\end{equation}
Finally, the temperature fluctuation $\Psi(x)$ satisfies
\begin{equation}\label{PsiBC}
	\Psi(0) = -\Psi(L) = \frac{\Delta T}{2}.
\end{equation} 
[$\Psi$ and $\Phi$ were introduced in Eq.~(\ref{TmuExp}).]\\

\subsection{Solution to the linearized hydrodynamic equations}

Employing standard thermodynamic identities and linearizing in $\Delta T$,
we rewrite the hydrodynamic 
Eqs.~(\ref{EntropyBalance})--(\ref{NumberImbalance}), (\ref{LinearResponse1}),  
and (\ref{LinearResponse2}) 
as the following set of 5 first order differential equations, 
valid to the lowest order in $u^2/v_{F}^2$:
\begin{align}
	\frac{1}{\bar{T}}\ddx{\Psi(x)}
	=&
	\frac{e^2}{6 \Pres \Dsig}\left[\NI J_{n}(x) - 2 u N_{\sigma}^2 \right]
	- \frac{u}{\vf \lel},
	\label{dPsi}\\
	\ddx{V_{\Phi}(x)}
	=&
	\frac{\bar{\mu}}{e\bar{T}}\ddx{\Psi(x)}
	\nonumber\\
	&+
	\frac{e}{2 \Dsig}
	\left[
	\frac{(\Sig{e}-\Sig{h})}{2}J_{n}(x) 
	+ \NII u
	\right],
	\label{dVPhi}\\
	\ddx{\mu_{I}(x)}
	=&
	\frac{e^2}{2 \Dsig}
	\left[
	-\frac{(\Sig{e}+\Sig{h})}{2}J_{n}(x) 
	+ \NI u
	\right],
	\label{dmuI}\\
	\ddx{J_{n}(x)}
	=&
	- \frac{2 n \lambda_{Q}}{\hbar} \mu_{I}(x),
	\label{dJn}\\
	\ddx{u} &= 0.
	\label{du}
\end{align}
The various new parameters appearing in Eqs.~(\ref{dPsi})--(\ref{du}) are defined by 
\begin{gather}
	\Sig{e} \equiv \sig{ee} - \sig{eh},\quad
	\Sig{h} \equiv \sig{hh} - \sig{eh},
	\nonumber\\
	\BSig{e} \equiv \sig{ee} + \sig{eh},\quad
	\BSig{h} \equiv \sig{hh} + \sig{eh},
	\nonumber\\
	\Dsig \equiv \sig{ee} \sig{hh} - \sig{eh}^2,
	\nonumber\\
	\NI
	\equiv
	\left(n_{e} \Sig{h} + n_{h} \Sig{e}\right),\quad	
	\NII
	\equiv
	\left(n_{e} \BSig{h} - n_{h} \BSig{e}\right),
	\nonumber\\
	N_{\sigma}^2 
	\equiv
	\left(
	n_{e}^2 \sig{hh} + n_{h}^2 \sig{ee} - 2 n_{e} n_{h} \sigma_{eh}
	\right).
	\label{Parameters}
\end{gather}
We note that the kinetic coefficient sum
\begin{equation}\label{Sige+Sigh}
	\Sig{e} + \Sig{h} = \sigin,
\end{equation}
the minimum conductivity at the Dirac point [Eqs.~(\ref{sigmaBulk}) and (\ref{siginDef})].

When supplemented with Eqs.~(\ref{BCmuI}) and (\ref{PsiBC}),  
Eqs.~(\ref{dPsi})--(\ref{du}) are easily solved. 
The results are
\begin{widetext}
\begin{align}\label{PsiSol}
	\Psi(x) 
	= 
	\frac{\Xi_{L} \Delta T}{2}
	&
	\left[ 
	\rc
	\frac{\NI^2}{\Dsig}
	\frac{2 \lQ}{L} 
	\frac{\sinh\left(\frac{L-2x}{2\lQ}\right)}{\cosh\left(\frac{L}{2\lQ}\right)}
	+
	\frac{\RR_{L} \, \sigin}{\kapinf} 
	\frac{(3 \Pres)^2}{\bar{T} e^2}
	\left(\frac{L-2x}{L}\right)
	\right],
\end{align}
\begin{align}\label{VPhiSol}
	V_{\Phi}(x)-V_{\Phi}(0)
	&=
	\frac{\bar{\mu}}{e\bar{T}}\Psi(x)
	+
	\left(\frac{\Delta T}{\bar{T}}\right)
	\frac{3 \Pres \Xi_{L}}{2 e}
	\left[
	\RR_{L}  
	\frac{\rho}{e}
	\left(\frac{2 x - L}{L}\right)
	+
	\rc
	\frac{\NI (\Sig{e}-\Sig{h})}{2 \Dsig}
	\frac{2 \lQ}{ L}
	\frac{\sinh\left(\frac{L-2x}{2\lQ}\right)}{\cosh\left(\frac{L}{2\lQ}\right)}
	\right],
\end{align}
\end{widetext}
\begin{align}\label{muISol}
	\mu_{I}(x)
	=
	-
	\left(\frac{\Delta T}{\bar{T}}\right)
	\rc \, \sigin
	&\frac{3 \Pres \NI \Xi_{L}}
	{4 \Dsig}
	\frac{2 \lQ}{L}
	\frac{\sinh\left(\frac{L-2x}{2\lQ}\right)}{\cosh\left(\frac{L}{2\lQ}\right)},
\end{align}
\begin{equation}\label{JnSol}
	J_{n}(x) = 
	\left(\frac{\Delta T}{\bar{T}}\right)
	\frac{6 \Pres \NI \Xi_{L}}{e^2 L}
	\left[
	\RR_{L} 
	-
	\rc 
	\frac{\cosh\left(\frac{L-2x}{2\lQ}\right)}{\cosh\left(\frac{L}{2\lQ}\right)}
	\right],
\end{equation}
\begin{align}\label{uSol}
	u
	&=
	\left(\frac{\Delta T}{\bar{T} }\right)
	\RR_{L}\,\sigin 
	\frac{3 \Pres}{L e^2}
	\Xi_{L},
\end{align}
where $\lQ$ denotes the imbalance relaxation length, defined via
\begin{equation}\label{lQDef}
	\frac{1}{\lQ^2}
	\equiv
	\frac{n \lambda_{Q} \sigin}
	{2 \Dsig} \frac{e^2}{\hbar}.
\end{equation}
In Eq.~(\ref{PsiSol}), $\kapinf$ is the thermal conductivity in 
the limit of infinite imbalance relaxation, as given by 
Eq.~(\ref{TCBulk}).
(See also Sec.~\ref{Sec: TCDiracPoint}, below.) 
Finally, the parameters $\RR_{L}$ and $\Xi_{L}$ appearing 
in Eqs.~(\ref{PsiSol})--(\ref{uSol}) are defined as
\begin{equation}\label{RRDef}
	\RR_{L}
	\equiv
	\rc +
	\frac{\hbar}{2 n e^2 \lambda_{Q} \lQ} \tanh\left(\frac{L}{2\lQ}\right),
\end{equation}
\begin{equation}\label{XiDef}
	\Xi_{L}
	\equiv
	\left[
	\rc \frac{\NI^2}{\Dsig}
	\frac{2 \lQ}{L} \tanh\left(\frac{L}{2\lQ}\right)
	+
	\frac{\RR_{L} \, \sigin}{\kapinf}
	\frac{(3 \Pres)^2 }{\bar{T} e^2}
	\right]^{-1}.
\end{equation}

In the next section, we specialize to the undoped case, and compute the
thermal conductivity $\kappa$ in the limit of infinite contact surface resistivity 
($\rc \rightarrow \infty$). We will discuss in detail the inhomogeneity of the temperature
and number current density profiles implied by Eqs.~(\ref{PsiSol}) and (\ref{JnSol}).
General results for both $\kappa$ and the thermopower $\TEP$ are obtained and 
discussed in the final Sec.~\ref{Sec: TCandTEP}.

%%%%%%%%%%%%%%%%%%%%%%%%%%%%%%%%%%%%%%%%%%%%%%%%%%%%%%%%%%%%%%%%%%%%%%%%%%%%%%%%%%%%%%%%%%%%%%%%%%%%%%
%%%%%%%%%%%%%%%%%%%%%%%%%%%%%%%%%%%%%%%%%%%%%%%%%%%%%%%%%%%%%%%%%%%%%%%%%%%%%%%%%%%%%%%%%%%%%%%%%%%%%%
%%%%%%%%%%%%%%%%%%%%%%%%%%%%%%%%%%%%%%%%%%%%%%%%%%%%%%%%%%%%%%%%%%%%%%%%%%%%%%%%%%%%%%%%%%%%%%%%%%%%%%

\section{Thermal conductivity at the Dirac point \label{Sec: TCDiracPoint}}

We turn now to the calculation of the kinetic coefficients characterizing
the thermoelectric transport. In this section, we concentrate upon the simplest 
case, that of zero doping. In equilibrium, the Dirac point is characterized by 
the conditions
\begin{gather}
	n_{e} = n_{h} = \frac{n}{2},\quad
	\bar{\mu} = 0,\nonumber\\
	\sig{ee} = \sig{hh}.
	\label{DiracPoint}
\end{gather}
As a result of the particle-hole symmetry implied by Eq.~(\ref{DiracPoint}),
the thermopower vanishes, $\TEP = 0$. [This result is demonstrated explicitly 
via Eq.~(\ref{TEPSol}) in Sec.~\ref{Sec: TCandTEP}].

The thermal conductivity $\kappa$ obtains from the heat current [Eq.~(\ref{JqDef})].
Within linear response, 
\begin{align}\label{HeatKappa}
	J_{q}
	&\equiv
	\kappa \frac{\Delta T}{L}
	\nonumber\\
	&=
	3 \Pres u
	+ \ord{\Delta T}^2. 
\end{align}
In this section, we assume an \emph{ideal} measurement of $\kappa$, in which the
contact electrical resistivity becomes arbitrarily large,
\begin{equation}\label{IdealContacts}
	\rc \rightarrow \infty
\end{equation}
[See Eqs.~(\ref{CurrentContBC1})--(\ref{CurrentContBC4}).]	
Then, we use Eq.~(\ref{uSol}) and impose in addition
the conditions listed in Eq.~(\ref{DiracPoint}) upon all equilibrium thermodynamic 
variables, arriving at the result 
\begin{align}
	\kappa
	= 
	\frac{
	\left(\frac{3 \mathcal{P}}{n}\right)^2 
	\frac{\sigma_{T}}{\bar{T} e^2}
	\frac{L}{2 l_{Q}}
	\coth\left(\frac{L}{2 l_{Q}}\right)
	}
	{
	1 
	+
	\frac{3 \mathcal{P} \sigma_{T}}{e^2 v_{F} n^2 l_{\mathsf{el}}}
	\frac{L}{2 l_{Q}}
	\coth\left(\frac{L}{2 l_{Q}}\right)
	},\label{TCDiracPoint}
\end{align}
where we have introduced 
\begin{align}\label{sigmaTDef}
	\sig{T} \equiv \sig{ee} + \sig{hh} + 2 \sig{eh}
	= 2\left(\sig{ee} + \sig{eh}\right).
\end{align}
As the so-defined $\kappa$ depends upon the sample length $L$, it is
sometimes more natural to introduce the thermal conductance $\Gth$,
\begin{equation}\label{GthDef}
	\Gth \equiv \frac{W}{L} \kappa,
\end{equation}
where $W$ is the width of the graphene slab perpendicular to the applied
thermal gradient.

The thermal conductance in Eqs.~(\ref{GthDef}) and (\ref{TCDiracPoint}) constitutes a
primary result of this paper. The imbalance relaxation due to non-electron and hole 
number-conserving inelastic collisions enters through the ratio $L/l_{Q}$, where the length 
$l_{Q}$ was defined by Eq.~(\ref{lQDef}). At the Dirac point, the latter simplifies to
\begin{equation}\label{lQDiracPoint}
	\frac{1}{\lQ^2}
	=
	\frac{2 n \lambda_{Q}}
	{\sig{T}} \frac{e^2}{\hbar}.
\end{equation}
The effects of quenched disorder are encoded in Eq.~(\ref{TCDiracPoint})
through the elastic mean free path $l_{\mathsf{el}}$.

\begin{figure}[t]
\includegraphics[width=0.45\textwidth]{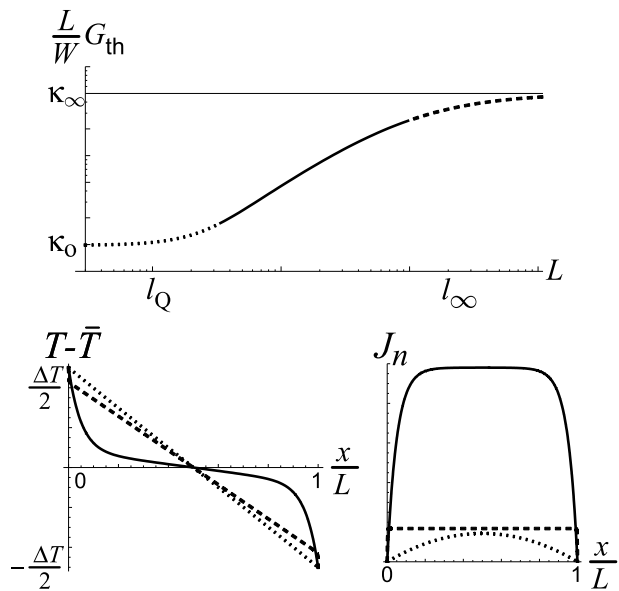}
\caption{
The top graph 
(a log-log plot)
depicts the qualitative form of the thermal ``conductivity'' 
$\kappa \equiv L G_{\mathsf{th}}/W$ [Eq.~(\ref{TCDiracPoint})] versus sample 
length $L$ for an undoped graphene strip possessing both slow imbalance relaxation 
and weak quenched disorder. Three manners of functional behavior for $\kappa$ are demarcated 
with dotted, solid, and dashed line segments; for each of these, the bottom plots 
show representative spatial profiles of the temperature $T(x) - \bar{T} = \Psi(x)$ 
[Eq.~(\ref{PsiSol})] and number current density $J_{n}(x)$ [Eq.~(\ref{JnSol})]. For system sizes 
$l_{Q} \lesssim L \lesssim \linf$ (solid curves), $\kappa$ grows linearly 
with $L$, while the temperature profile is inhomogeneous; this regime is also characterized by
the maximal particle flux $J_{n}$, as measured at the device center $x = L/2$. 
By contrast, in the short ($L \ll l_{Q}$) and long ($L \gg \linf$) sample size limits,
respectively designated by dotted and dashed curves in the figure, $\kappa$ saturates 
to $\kappa_{\circ}$ and $\kapinf$, respectively; here, the temperature profile asymptotes 
to a linear gradient. The assumption of infinite contact resistivity [Eq.~(\ref{IdealContacts})]
ensures that $J_{n}(x)$ vanishes at $x = 0,L$.
\label{TempCurrentDensity}}
\end{figure}

Let us interpret our results. Consider first the clean limit, with $l_{\mathsf{el}} \rightarrow \infty$.
In this case, the physics is determined by the ratio of the system size $L$ to the length scale $l_{Q}$ 
[Eq.~(\ref{lQDiracPoint})]. 
For a short device, $L \ll l_{Q}$, the thermal conductivity given by Eq.~(\ref{TCDiracPoint})
asymptotes to
\begin{align}\label{TCDiracPointShortClean}
	\kappa 
	&=
	\kappa_{\circ}
	+ \ord{\frac{L}{l_{Q}}}^2,
	\nonumber\\ 
	\kappa_{\circ}
	&\equiv
	\left[\frac{3^3 \zeta(3)}{\pi^2}\right]^2 \frac{k_{B}^2 \bar{T}}{e^2} \sigma_{T}.
\end{align}
The prefactor in this last equation obtains from the equilibrium pressure and density 
in Eq.~(\ref{TCDiracPoint}), evaluated for the ideal relativistic quantum gas at zero 
doping, taking into account valley and spin degeneracies in graphene. In this short device limit, 
the temperature profile $\Psi(x)$ [Eq.~(\ref{PsiSol})] falls approximately linearly
across the entire sample, while the number current density $J_{n}$ is small everywhere along the strip.

In the opposite limit of a long device, $L \gg l_{Q}$, the thermal conductance defined via
Eq.~(\ref{GthDef}) approaches the $L$-independent constant
\begin{equation}\label{GthDiracPointLongClean}
	\Gth = \frac{W}{2 l_{Q}} \kappa_{\circ}
	+ 
	\ordSQ{
	\exp\left(-\frac{L}{\lQ}\right)
	}.
\end{equation}
The temperature profile $T(x) = \bar{T} + \Psi(x)$ now consists of three regions: within a distance
$l_{Q}$ of the sample boundaries, the temperature drops approximately linearly; in between these boundary
regions, $T \sim \bar{T}$. A large carrier number current $J_{n}(x)$ [Eq.~(\ref{JnSol})] flows through 
the bulk of the sample, but pinches off to zero at $x = 0$ and $x = L$ where, for $\Delta T > 0$, 
generation and recombination processes, respectively, relax the accumulating population imbalance. 
As indicated by Eqs.~(\ref{JnSol}) and (\ref{RRDef}), if we relax the condition stipulating ideal contacts 
[Eq.~(\ref{IdealContacts})] by assuming a non-zero contact conductance density $1/\rc > 0$, then 
$J_{n}(x)$ adopts a non-zero value at the sample edges; in this situation, electrons and holes 
that penetrate (escape) the graphene are generated (recombined) in the contacts.

For a long device ($L \gg l_{Q}$) possessing, in addition, weak quenched disorder, the graphene sample 
behaves as three thermal resistors in series. The boundary resistance is dominated by the imbalance relaxation 
processes as in Eq.~(\ref{GthDiracPointLongClean}), but there is now a finite temperature drop through the bulk.
Disorder also introduces a second length scale $\linf \gg l_{Q}$ into the denominator of Eq.~(\ref{TCDiracPoint}), 
defined as
\begin{equation}\label{DisorderScale}
	\linf \equiv \lel \sqrt{\frac{e^2}{\hbar \sigma_{T} \lambda_{Q}}}.
\end{equation}
The scale $\linf$ emerges from Eq.~(\ref{TCDiracPoint}) when all thermodynamic variables
in that equation are evaluated for the ideal quantum relativistic gas with zero charge density. 
For a graphene slab with $L \gg \linf$, the bulk thermal resistance dominates, and the thermal 
conductivity $\kappa$ asymptotes to the infinite imbalance relaxation value given by Eq.~(\ref{TCBulk}), 
which simplifies at the Dirac point to
\begin{equation}\label{DisorderLimitedKappa}
	\kapinf = \frac{3 \mathcal{P} v_{F} \lel}{\bar{T}}.
\end{equation}
For an appropriate definition of $l_{\mathsf{el}}$, Eq.~(\ref{DisorderLimitedKappa}) is
the same result obtained in previous work,\cite{SachdevGrapheneHYDRO} which effectively
assumed infinite imbalance relaxation. In Fig.~\ref{TempCurrentDensity}, we sketch our results for the
$L$-dependence of the temperature and number current profiles $T(x)$ and $J_{n}(x)$, as well as the 
thermal conductivity $\kappa = L \Gth/W$, for a sample satisfying $\linf \gg \lQ$. 

Note that in the clean limit $\lel \rightarrow \infty$, the result in Eq.~(\ref{DisorderLimitedKappa}) 
appears to suggest that the thermal conductivity at zero doping diverges for any $L$. For any finite 
imbalance relaxation $\lQ > 0$, we have seen that this conclusion is incorrect; instead, the response
is inhomogeneous (Fig.~\ref{TempCurrentDensity}), yielding a finite $\Gth$ for all $L$ 
[Eqs.~(\ref{GthDef}) and (\ref{TCDiracPoint})]. We observe that $\kappa$ in
Eq.~(\ref{TCDiracPoint}) depends only upon the combination $\sigma_{T}$ [Eq.~(\ref{sigmaTDef})]
of ``intrinsic'' kinetic coefficients; 
this is different from the combination that enters
into the minimum conductivity at the Dirac point $\sigin$ [Eqs.~(\ref{sigmaBulk}) and 
(\ref{siginDef})].
Finally, it is important to stress that the limit given by Eq.~(\ref{DisorderLimitedKappa}) is 
appropriate only to the hydrodynamic transport regime, $\tauel \gtrsim \tauin$; in the opposite 
case, $\kappa$ is constrained by the ``generalized'' Wiedemann-Franz law,\cite{footnote-b}
and a different result can be obtained.

%%%%%%%%%%%%%%%%%%%%%%%%%%%%%%%%%%%%%%%%%%%%%%%%%%%%%%%%%%%%%%%%%%%%%%%%%%%%%%%%%%%%%%%%%%%%%%%%%%%%%%
%%%%%%%%%%%%%%%%%%%%%%%%%%%%%%%%%%%%%%%%%%%%%%%%%%%%%%%%%%%%%%%%%%%%%%%%%%%%%%%%%%%%%%%%%%%%%%%%%%%%%%
%%%%%%%%%%%%%%%%%%%%%%%%%%%%%%%%%%%%%%%%%%%%%%%%%%%%%%%%%%%%%%%%%%%%%%%%%%%%%%%%%%%%%%%%%%%%%%%%%%%%%%

\section{Thermoelectric transport coefficients: Arbitrary doping \label{Sec: TCandTEP}}

In this section, we compute the thermoelectric response of graphene at non-zero doping.
Combining Eqs.~(\ref{HeatKappa}) and (\ref{uSol}), the general
expression for the thermal conductivity is
\begin{align}\label{TCSol}
	\kappa
	&=
	\RR_{L} \, \sigin
	\frac{(3 \Pres)^2}{\bar{T} e^2}
	\Xi_{L}.
\end{align}

The thermoelectric power $\TEP$ was identified in Sec.~\ref{Sec: Hydro} as the ratio
expressed in Eq.~(\ref{TEPDef}), where $\Delta V_{w} = \dVwL - \dVwO$ is the voltage 
drop across the ``wire'' interconnect serving as our voltmeter in the experiment
sketched in Fig.~\ref{FigSetup}. The boundary conditions in Eq.~(\ref{BCdVwL})
demonstrate that $\TEP$ is determined by the electrochemical
potential drop $\Delta V_{\Phi} = V_{\Phi}(L) - V_{\Phi}(0)$ across the graphene. 
Eq.~(\ref{VPhiSol}) then implies that
\begin{align}\label{TEPSol}
	\TEP 
	=&
	\frac{\Delta V_{\Phi}}{\Delta T}
	\nonumber\\
	=&
	-\frac{\bar{\mu}}{\bar{T} e}
	+
	\RR_{L}
	\frac{3 \Pres \Xi_{L}}{\bar{T} e^2}
	\rho  
	\nonumber\\
	&-
	\rc
	\frac{3 \Pres \Xi_{L}}{\bar{T} e}
	\frac{\NI (\Sig{e}-\Sig{h})}{2 \Dsig} 
	\frac{2 \lQ}{L}
	\tanh\left(\frac{L}{2\lQ}\right).
\end{align}
The parameters $\RR_{L}$, $\Xi_{L}$, $\NI$, $\Sig{e,h}$, and $\Dsig$ appearing
in Eqs.~(\ref{TCSol}) and (\ref{TEPSol}) were defined in Eqs.~(\ref{Parameters}), 
(\ref{RRDef}), and (\ref{XiDef}), above.

The general expressions in Eqs.~(\ref{TCSol}) and (\ref{TEPSol}) 
constitute the primary results of this paper. We now specialize these 
results to the short ($L \ll \lQ$) and long ($L \gg \lQ$) device limits.

\subsubsection{Short device $L \ll \lQ$}

For a device shorter than the imbalance relaxation length $\lQ$ [Eq.~(\ref{lQDef})], 
Eqs.~(\ref{TCSol}) and (\ref{TEPSol}) asymptote to
\begin{align}
	\kappa 
	=&
	\frac{(3 \Pres)^2}{\bar{T} e^2}
	\left[
	\frac{
	8 \rc \Dsig
	+
	L
	\sigin
	}
	{
	8 \rc
	\left(
	N_{\sigma}^2
	+
	\frac{3 \Pres \Dsig}{e^2 \vf \lel} 
	\right)
	+
	L
	\frac{(3 \Pres)^2 \sigin}{\bar{T} e^2 \kapinf}
	}
	\right]
	\nonumber\\
	&+ \ord{\frac{\lQ}{L}},
	\label{TCshortFinal}
\end{align}
\begin{align}
	\TEP
	=&
	\frac{3 \Pres}{\bar{T} e^2}
	\left[
	\frac{
	4\rc e \NII
	+
	L
	\rho
	}
	{
	8 \rc
	\left(
	N_{\sigma}^2
	+
	\frac{3 \Pres \Dsig}{e^2 \vf \lel} 
	\right)
	+
	L
	\frac{(3 \Pres)^2 \sigin}{\bar{T} e^2 \kapinf}
	}
	\right]
	\nonumber\\
	&-\frac{\bar{\mu}}{\bar{T} e}
	+ \ord{\frac{\lQ}{L}}.
	\label{TEPshortFinal}
\end{align}
These results indicate that the limit $L \ll \lQ$ gives not one but several regimes 
for the thermoelectric response in the general case; the demarcation lines between these
depend upon $L$ and the contact surface resistivity $\rc$, as well as the
doping and the extent of disorder in the sample. 

We analyze Eqs.~(\ref{TCshortFinal}) and (\ref{TEPshortFinal}), neglecting the influence
of disorder for simplicity ($\lel \rightarrow \infty$). 
We focus upon the non-degenerate case of high temperatures and low doping, 
$k_{B} \bar{T} \gg |\bar{\mu}|$. 
A maximum of three behavioral regimes are possible for both 
$\kappa$ and $\TEP$, and these are accessed sequentially with increasing $L$.   
We define two length scales mediated by the contact resistivity $\rc$,
\begin{equation}\label{lnumldenDef}
	\lnum \equiv \rc \, \sigin,\quad
	\lden \equiv \rc \, \sigin \left(\frac{e\,n}{\rho}\right)^2,
\end{equation}
where $n$ and $\rho$ are the total number and charge densities, respectively [Eq.~(\ref{nrhoDef})].
In Eq.~(\ref{lnumldenDef}), $\sigin$ denotes the minimum dc conductivity at the
Dirac point in the hydrodynamic regime, as defined by Eqs.~(\ref{sigmaBulk}) and
(\ref{siginDef}). Clearly we have $\lnum \ll \lden$ in the non-degenerate regime.
The interpretation of $\lnumden$ is as follows. At the scale $\lnum$, the electrical
conductance of the graphene sample \emph{at zero doping} is of order the contact conductance, since
\begin{align}\label{lnumInterp}
	\frac{W \sigin}{\lnum} \sim \frac{1}{R_{c}}
\end{align} 
where $R_{c} = \rc/W$, and $W$ is the sample width transverse to the current flow.
By contrast, $\lden$ is the scale at which the thermal conductance $\Gth$ for clean,
non-degenerate graphene in the infinite imbalance relaxation limit becomes of order 
the electrical contact resistance:
\begin{align}\label{ldenInterp}
	G_{\mathsf{th},\infty}\left(L = \lden \right) 
	&\equiv
	\frac{W \kapinf}{\lden}
	\sim
	\left(\frac{k_{B}^2 \bar{T}}{e^2}\right) \frac{1}{R_{c}}, 
\end{align}
where $\kapinf$ is given by Eqs.~(\ref{TCBulk}) and (\ref{sigmaBulk}) in the limit $\lel \rightarrow \infty$.

For a device with $L \ll \lnum$, we can set $L = 0$ everywhere in both Eqs.~(\ref{TCshortFinal})
and (\ref{TEPshortFinal}). The resulting expressions for $\kappa$ and $\TEP$ are exactly
those obtained for graphene possessing zero imbalance relaxation, $\lambda_{Q} = 0$ in 
Eqs.~(\ref{LinearResponse2}) and (\ref{lQDef}), as measured through
``ideal'' (electrically insulating, thermally conducting) contacts. 
These zero imbalance relaxation (shortest device) expressions for $\kappa$ and $\alpha$
are completely different from the infinite imbalance relaxation (longest device) results
$\kapinf$ and $\TEPinf$ given by Eqs.~(\ref{TCBulk}) and (\ref{TEPBulk}), above. 
For example, using the definitions of $\NII$ and $N_{\sigma}^2$ from Eq.~(\ref{Parameters}),
it is clear from Eq.~(\ref{TEPshortFinal}) that the thermopower in this regime vanishes smoothly 
as the sample charge density is tuned through the Dirac point, even for a perfectly clean sample.
By contrast, Eq.~(\ref{TEPBulk}) gives $\TEPinf = s/\rho$ in the limit $\lel \rightarrow \infty$,
which exhibits a simple pole at $\rho = 0$ ($s$ is the entropy density).  
Measurements of the thermoelectric response in both the short and long device limits for graphene
in the hydrodynamic regime should allow extraction of the independent diagonal ($\sig{ee}$) and 
off-diagonal ($\sig{eh}$) tensor coefficients defined by Eq.~(\ref{LinearResponse2}).

In a longer device with intermediate $L$, $\lnum \ll L \ll \lden$, the terms proportional
to $L$ in the numerators of both Eqs.~(\ref{TCshortFinal}) and (\ref{TEPshortFinal}) dominate 
the response. Similar to the intermediate regime of the thermal conductivity of undoped, 
disordered graphene discussed in Sec.~\ref{Sec: TCDiracPoint} and illustrated in 
Fig.~\ref{TempCurrentDensity}, both $\kappa$ and $\TEP$ rise linearly with $L$.
For much longer devices with $L \gg \lden$, the term proportional to $L$ in the
denominator of both Eqs.~(\ref{TCshortFinal}) and (\ref{TEPshortFinal}) dominates, so that
$\kappa$ and $\TEP$ plateau to their infinite imbalance relaxation limits
($\kapinf,\TEPinf$), transcribed above in Eqs.~(\ref{TCBulk}) and (\ref{TEPBulk}).
In other words, when the contact electrical conductance becomes larger than the 
infinite imbalance relaxation limit of the graphene thermal conductance 
$G_{\mathsf{th},\infty}$ (times $e^2/k_{B}^2 \bar{T}$) [Eq.~(\ref{ldenInterp})], 
we find that $\kappa \rightarrow \kapinf$.

Observation of the described crossovers requires that the imbalance relaxation length $\lQ$ 
exceeds one or both of $\lnum \ll \lden$. In the non-degenerate regime of the carrier plasma, 
we obtain an order of magnitude estimate for $\lQ$ by approximating Eq.~(\ref{lQDef}) as
\begin{equation}\label{lQNonDeg}
	\lQ \sim \sqrt{\frac{\hbar \, \sigin}{e^2 \lambda_{Q}}} \frac{\hbar \vf}{k_{B} \bar{T}}.
\end{equation}
[See also Eq.~(\ref{lQDiracPoint}), Sec.~\ref{Sec: TCDiracPoint}].
For a fixed sample size $L$, we can determine the temperatures $T \equiv \Trsigkap$ below
which the crossovers at $L \sim \lnum$ and $L \sim \lden$ become observable within the short device
($L \ll \lQ$) regime.
As it is expected to vary only weakly with decreasing temperature, the smaller scale $\lnum$ may be
approximated as a constant. Eq.~(\ref{lQNonDeg}) then implies that the crossover at $L \sim \lnum$ 
occurs for temperatures
\begin{equation}\label{TrsigDef}
	\bar{T} \lesssim 
	\Trsig
	\equiv
	\frac{\hbar \vf}{k_{B}}
	\frac{1}{\rc}
	\sqrt{\frac{\hbar}{e^2 \sigin \lambda_{Q}}}.
\end{equation} 
By contrast, the $\bar{T}$-dependence of $\lden$ is partially determined by the
conditions of the experiment. If for example the charge density $\rho = e(n_{e} - n_{h})$ is held 
constant as the temperature is varied, then Eq.~(\ref{lQNonDeg}) suggests that the crossover 
at $L \sim \lden$ should be observable in the short device regime only for temperatures 
\begin{equation}\label{TrkapDef}
	\bar{T} \lesssim 
	\Trkap
	\equiv
	\frac{\hbar \vf}{k_{B}} 
	\left(\frac{\rho^2}{e^2 \, \rc}\sqrt{\frac{\hbar}{e^2 \sigin \lambda_{Q}}}\right)^{1/5}.
\end{equation}
These equations hold only for relatively clean, non-degenerate graphene; the average chemical 
potential and temperature must satisfy $|\bar{\mu}|/k_{B} \ll \bar{T}$.
In addition, if we take $\lambda_{Q} \sim 1$ and $\sigin \sim 4 e^2/h$
(see the discussion in the next subsection, below), then Eq.~(\ref{TrkapDef})
holds only for $\lel \gg \lQ (e \, n/\rho)^2$. By contrast, for
$\lel \ll \lQ$, only one crossover within the $L \ll \lQ$ regime is possible,
at $L \sim \lnum$.

\subsubsection{Numbers for the short device ($L \ll \lQ)$ regime\label{Sec: Numbers}}

Experimentally, $\vf \sim 10^6$ m/s, while $\sigin \sim 4 e^2/h$.\cite{CastroNetoReview,GrapheneExpt} 
We have not calculated the imbalance relaxation parameter $\lambda_{Q}$ in 
Eqs.~(\ref{lQNonDeg})--(\ref{TrkapDef}), which requires a quantum kinetic equation treatment 
that incorporates three-particle collisions and/or impurity-assisted recombination. We have argued that, 
due to the absence of two-particle mechanisms (see Fig.~\ref{FigNonDecay} and the concomitant discussion 
in Sec.~\ref{Sec: Intro}), imbalance relaxation should be a slow process in the hydrodynamic regime. 
In what follows, we take the conservative estimate $\lambda_{Q} \sim 1$.

We consider first the crossover at $L \sim \lnum$. 
For a contact resistance of $R_{c} = 100 \, \Omega$ and a relative carrier density of 
$\rho/e = 10^{10}/\mathrm{cm}^2$, Eq.~(\ref{TrsigDef}) gives $\Trsig \sim 400 K$. The 
average (relative) chemical potential is $\bar{\mu}/k_{B} \sim 20 K$ for the assumed 
density,\cite{footnote-d} while $\lnum \sim \lQ(\Trsig) \sim 0.02 \, \mu\mathrm{m}$. 
Longer crossover lengths $\lnum$ and lower temperatures $\Trsig$ can be obtained for 
larger contact resistances $R_{c}$, but lower carrier densities are required to preserve 
the condition of non-degeneracy for the carrier plasma. A contact resistance of 
$R_{c} = 10 \, k\Omega$ and a relative carrier density of $\rho/e = 10^{6}/\mathrm{cm}^2$ 
gives $\Trsig \sim 4 K$, with $\lnum \sim \lQ(\Trsig) \sim 2 \, \mu\mathrm{m}$. In this 
case, $\bar{\mu}/k_{B} \sim 0.2 K$.

We now turn to the crossover at $L \sim \lden$.
For a contact resistance $R_{c} = 1 \, \Omega$ and a relative density $\rho/e = 10^{11}/\mathrm{cm}^2$, 
Eq.~(\ref{TrkapDef}) gives $\Trkap \sim 700 K$. The chemical potential is $\bar{\mu}/k_{B} \sim 100 K$ for
the assumed density,\cite{footnote-d} while $\lden \sim \lQ(\Trkap) \sim 0.01 \, \mu\mathrm{m}$. 
In order to observe deviations from the infinite imbalance relaxation limit 
[Eqs.~(\ref{sigmaBulk})--(\ref{TEPBulk})] in the short device regime,
the sample length $L \lesssim \lden$, so these conditions would seem to require an 
impractically short device. Longer devices meeting the required constraints are possible for more resistive
contacts and lower carrier densities. For a contact resistance of $R_{c} = 100 \, \Omega$ and a relative
density $\rho/e = 10^{7}/\mathrm{cm}^2$, one finds $\Trkap \sim 7 K$, $\bar{\mu}/k_{B} \sim 1 K$, and 
$\lden \sim \lQ(\Trkap) \sim 1 \, \mu\mathrm{m}$.  Contact resistance can be controlled in principle through the 
incorporation of a highly insulating spacer layer of varying thickness between the graphene and the contact metal.

\subsubsection{Long device $L \gg \lQ$}

In the opposite limit where the sample length $L$ exceeds the imbalance relaxation length $\lQ$, 
Eqs.~(\ref{TCSol}) and (\ref{TEPSol}) simplify as follows:
\begin{align}\label{TClongFinal}
	\kappa 
	=&
	\frac{\kapinf}
	{
	1+
	\left(\frac{\lQ}{L}\right)
	\frac{8 \, \rc \, \kapinf \, \bar{T} e^2 \NI^2}
	{
	(3 \Pres)^2 \sigin
	\left[
	4 \Dsig \rc+
	\lQ \sigin
	\right]
	}
	}
	\nonumber\\
	&+\ordSQ{\exp\left(-L/\lQ\right)},
\end{align}
\begin{align}\label{TEPlongFinal}
	\TEP
	=&
	\frac{\kappa \rho}{3 \Pres \sigin}
	-\frac{\bar{\mu}}{\bar{T} e}
	\nonumber\\
	&
	-
	\left(\frac{\lQ}{L}\right)
	\frac{e \kappa}{3 \Pres \sigin}
	\frac{4 \rc \NI (\Sig{e}-\Sig{h})}
	{\left[
	4 \Dsig \rc
	+
	\lQ
	\sigin
	\right]}
	\nonumber\\
	&+\ordSQ{\exp\left(-L/\lQ\right)}.
\end{align}

The denominator of Eq.~(\ref{TClongFinal}) introduces yet another length 
scale into the problem,
\begin{align}\label{linfDef2}
	\linf
	&\equiv
	\lQ
	\frac{8 \, \rc \, \kapinf \, \bar{T} e^2 \NI^2}
	{
	(3 \Pres)^2 \sigin
	\left[
	4 \Dsig \rc+
	\lQ \sigin
	\right]
	}
	\nonumber\\
	&\sim
	\frac{\lQ\,\lden}{\lQ + \lnum},
\end{align}
where $\lnumden$ were introduced in Eq.~(\ref{lnumldenDef}).
On the second line of Eq.~(\ref{linfDef2}), we have used Eq.~(\ref{TCBulk}), 
neglected the effects of disorder for simplicity, approximated
$\Sig{e} \sim \Sig{h} \sim \sigin$ and $\Dsig \sim \sigin^2$, 
and we have assigned to all thermodynamic variables their values at 
the Dirac point. In the limit $L \gg \linf$, Eqs.~(\ref{TClongFinal}) and
(\ref{TEPlongFinal}) show that the thermal transport coefficients
asymptote toward their infinite imbalance relaxation limits, 
$\kappa \rightarrow \kapinf$ and $\TEP(\kappa) \rightarrow \TEP(\kapinf) = \TEPinf$.

The criterion for the existence of a length-dependent crossover in the behavior
of $\kappa$ and $\TEP$ within the $L \gg \lQ$ regime is as follows. For $\lQ \gg \lnum$, 
Eq.~(\ref{linfDef2}) gives
\begin{equation}\label{linfCross1}
	\linf \sim \lden.
\end{equation}
The scale $\lden$ is also the crossover scale to the same (effective) infinite imbalance relaxation 
limit, as obtained in the \emph{opposite} regime $L \ll \lQ$. Thus, the location
of the crossover to infinite imbalance relaxation behavior relative to $L = \lQ$ 
depends upon the unspecified ratio $\lQ/\lden$, consistent with the previous discussion.

In the opposite limit $\lQ \ll \lnum$, a crossover in the $L \gg \lQ$ regime definitely
occurs at
\begin{equation}\label{linfCross2}
	\linf \sim \lQ \left(\frac{e \, n}{\rho}\right)^2 \ll \lden,
\end{equation}
with $n$ ($\rho$) the total number (charge) density. In this case, an intermediate
regime exists for $\lQ \ll L \ll \linf$, in which $\kappa$ and $\TEP$ grow linearly
with $L$. Only for $L \gg \linf$ do the infinite imbalance relaxation limits for
these kinetic coefficients emerge.

We stress once again that Eqs.~(\ref{TCshortFinal}), (\ref{TEPshortFinal}), (\ref{TClongFinal}),
and (\ref{TEPlongFinal}) hold only for the case of hydrodynamic, interparticle collision-mediated 
transport. In the opposite case of ``disorder-limited'' transport, where the elastic scattering
rate exceeds the inelastic rate due to interparticle collisions ($\tauel \lesssim \tauin$), 
$\kappa$ and $\TEP$ are slaved to the electrical conductivity through the 
``generalized'' Wiedemann-Franz law and Mott relation, respectively.\cite{footnote-b} 
Further discussion on the distinction between interaction and disorder-limited transport in 
graphene can be found in Ref.~\onlinecite{DGKP1}.

Finally, we comment upon the physics of the thermoelectric transport within the $L \gg \linf$ regime.
From Eq.~(\ref{muISol}), we note that, in the long device limit, the
imbalance chemical potential $\mu_{I}(x)$ [introduced in Eq.~(\ref{muDefs})] is exponentially
suppressed between boundary layers of size $\lQ$. The electron-hole population imbalance
is therefore confined to the boundary regions. By contrast, Eq.~(\ref{JnSol}) shows that
the number current $J_{n}$ that flows through the bulk of the sample decays only linearly with 
increasing $L$, for fixed $\Delta T$: 
\begin{align}
	J_{n}(L/2) 
	&\stackrel[L \gg \linf]{}{\sim} 
	\kapinf
	\Delta T
	\frac{2 \NI}{3 \Pres L \sigin}
	\nonumber\\
	&\sim
	\frac{\kapinf}{L}
	\frac{\Delta T}{k_{B} \bar{T}}
	\label{Jlong},
\end{align}
where we have used Eq.~(\ref{TCSol}) and approximated $\Sig{e} \sim \Sig{h}$.
For $L \gg \lQ$, $\kappa \propto L$ at the Dirac point for clean graphene [Eq.~(\ref{TCDiracPoint})
with $\lel \rightarrow \infty$]; at zero doping $J_{n}$ therefore saturates to a finite, non-zero 
value as the system size diverges, consistent with the picture of the central region as a ``perfectly 
conducting thermal wire.'' By contrast, the $1/L$ decay of Eq.~(\ref{Jlong}) away from the Dirac 
point is consistent with the finite thermal drop implied by $\kapinf$ in Eqs.~(\ref{TClongFinal}) 
and (\ref{TCBulk}).

%%%%%%%%%%%%%%%%%%%%%%%%%%%%%%%%%%%%%%%%%%%%%%%%%%%%%%%%%%%%%%%%%%%%%%%%%%%%%%%%%%%%%%%%%%%%%%%%%%%%%%
%%%%%%%%%%%%%%%%%%%%%%%%%%%%%%%%%%%%%%%%%%%%%%%%%%%%%%%%%%%%%%%%%%%%%%%%%%%%%%%%%%%%%%%%%%%%%%%%%%%%%%
%%%%%%%%%%%%%%%%%%%%%%%%%%%%%%%%%%%%%%%%%%%%%%%%%%%%%%%%%%%%%%%%%%%%%%%%%%%%%%%%%%%%%%%%%%%%%%%%%%%%%%

\section{Conclusion \label{Sec: Conclusion}}

In summary, we have demonstrated that thermoelectric transport in graphene within the hydrodynamic 
regime exhibits a range of behaviors when the finite rate of carrier imbalance relaxation is
taken into account. Since the relativistic spectrum of clean graphene is non-decaying, the lowest
order two-particle recombination and generation processes are kinematically forbidden, suggesting
that the imbalance relaxation lifetime $\tau_{Q}$ might significantly exceed other intrinsic graphene
timescales. 

The essential transport physics in the hydrodynamic regime is encoded by four intrinsic parameters:
these are the minimum conductivity at the Dirac point $\sigin$, the off-diagonal (or ``drag'') conductivity 
$\sig{eh}$, the imbalance relaxation length $\lQ$, and the elastic mean free path $\lel$. 
Of these, $\sigin$, $\sig{eh}$, and $\lQ$ are mediated entirely by intercarrier collisions.
The parameters $\sigin$ and $\lel$ can be obtained from measurement of 
the bulk conductivity (at variable doping), or the combined measurement of the electronic thermal 
conductivity $\kappa(L)$ and the thermopower $\alpha(L)$, in the limit of a long device with $L \gg \lQ$.
The drag conductivity $\sig{eh}$ can be extracted from a measurement of either $\kappa$ or $\alpha$
in the opposite, short device limit $L \ll \lQ$. For a sample with $L \gtrsim \lQ$, a local probe of
either the electronic temperature or electrochemical potential profiles should allow determination
of $\lQ$, since these are predicted to be inhomogeneous, with boundary layers of size $\lQ$ confined
near the device terminals.

We have given general formulae for both $\kappa$ and $\alpha$ at arbitrary doping and device size $L$, 
incorporating the effects of non-ideal contacts. Non-ideal contacts allow exchange of carriers with the 
graphene, providing an alternate route for imbalance relaxation. We have explicated the various crossover 
regimes that separate the zero imbalance relaxation (short device) and infinite relaxation (long device) 
limiting behaviors.

In this paper, we have neglected the effects of phonons in graphene. 
Both acoustic and optical phonons can influence the electronic thermal conductivity 
contribution $\kappa$ and the thermopower $\alpha$, through inelastic electron-phonon scattering.
Specifically, real electron-phonon collisions may modify (i) the imbalance relaxation rate due to 
electron-hole pair to phonon conversion processes, (ii) the inhomogeneous electronic temperature 
profile, due to energy exchange with the phonon bath, and (iii) the thermoelectric power through phonon 
drag.
By contrast, virtual electron-phonon interactions are strongly irrelevant, and 
the concomitant renormalization effects may be typically neglected.

For temperatures less than $\hbar \omega_{\mathsf{ph}}/k_{B} \sim 700 K$, all optical modes are 
frozen-out.\cite{MounetMarzari} Untethered (``free floating'') graphene supports linearly-dispersing 
acoustic phonons within the transverse (TA) and longitudinal (LA) in-plane modes, as well as
quadratically-dispersing phonons in an out-of-plane (ZA) mode. 
Under tension imposed by external contacts or surface 
adhesion to a substrate, however, the ZA dispersion also becomes linear.\cite{CastroNetoReview}
 
Because the acoustic phonon velocities\cite{OnoSugihara} $\vph \sim 10^4$ m/s $\ll \vf$,
the electron-hole pair creation and annihilation processes  
\[
	e^{-} + h^{+} \leftrightarrow ph
\]
are kinematically forbidden.
Therefore the electron-phonon scattering does not contribute to the imbalance relaxation, at least
to lowest order. For graphene in the non-degenerate regime with $k_{B} T \ll \hbar \omega_{\mathsf{ph}}$,
the in-plane acoustic phonon contribution to the inelastic electron lifetime (due to electron and hole number-conserving
processes) $\tauph \propto (k_{B} T)^{-2}$, by dimensional analysis. The associated electron-phonon relaxation 
length $\lph \equiv \vf \tauph$ can be estimated with the Boltzmann transport result\cite{e-phLifetime}
\begin{equation}\label{lphDef}
	\lph \sim \frac{4 (\hbar \vf)^3 \rho_{m} \vph^2}{\left(k_{B} T D\right)^2},
\end{equation}
where $\rho_{m} \sim 7.6\times 10^{-7}$ $\mathrm{kg/m}^2$ denotes the 2D mass density of graphene,
$\vph \sim 2 \times 10^{4}$ m/s is the phonon velocity for the LA mode, and $D \sim 19$ $\mathrm{eV}$ is 
the deformation potential.\cite{OnoSugihara} Using these parameters, 
$\lph \sim 80$ $\mu$m at $T = 100 K$ and $0.8$ cm at $T = 10 K$. For devices with sample dimensions 
$L,W \lesssim \lph$, phonons may be neglected. In larger devices, the loss of carrier energy to the 
phonon bath on scales longer than $\lph$ becomes important, so that the parameter $\lph$ will enter 
e.g.\ into the temperature profile across the device. In addition, phonon drag effects can become 
important in sufficiently clean samples with $L \gg \lph$.

\begin{acknowledgments}

We thank Yuri Zuev, Philip Kim, and Nadia Pervez for helpful discussions,
and Leonid Glazman and Leon Balents for reading the manuscript.
This work was supported in part by the Nanoscale Science and Engineering Initiative of the
National Science Foundation under NSF Award Number CHE-06-41523, and by the New York State Office 
of Science, Technology, and Academic Research (NYSTAR) (M.S.F.).

\end{acknowledgments}

%%%%%%%%%%%%%%%%%%%%%%%%%%%%%%%%%%%%%%%%%%%%%%%%%%%%%%%%%%%%%%%%%%%%%%%%%%%%
%%%%%%%%%%%%%%%%%%%%%%%%%%%%%%%%%%%%%%%%%%%%%%%%%%%%%%%%%%%%%%%%%%%%%%%%%%%%
%%%%%%%%%%%%%%%%%%%%%%%%%%%%%%%%%%%%%%%%%%%%%%%%%%%%%%%%%%%%%%%%%%%%%%%%%%%%
%%%%%%%%%%%%%%%%%%%%%%%%%%%%%%%%%%%%%%%%%%%%%%%%%%%%%%%%%%%%%%%%%%%%%%%%%%%%
%%%%%%%%%%%%%%%%%%%%%%%%%%%%%%%%%%%%%%%%%%%%%%%%%%%%%%%%%%%%%%%%%%%%%%%%%%%%
%%%%%%%%%%%%%%%%%%%%%%%%%%%%%%%%%%%%%%%%%%%%%%%%%%%%%%%%%%%%%%%%%%%%%%%%%%%%
%%%%%%%%%%%%%%%%%%%%%%%%%%%%%%%%%%%%%%%%%%%%%%%%%%%%%%%%%%%%%%%%%%%%%%%%%%%%


\begin{thebibliography}{99}

\bibitem{CastroNetoReview}
	For a recent review, see e.g.\
	A. H. Castro Neto, F. Guinea, N. M. R. Peres, K. S. Novoselov, and A. K. Geim,
	arXiv:0709.1163v2 [cond-mat.other] (2008).
\bibitem{NomuraMacDonald}
	K. Nomura and A. H. MacDonald, Phys. Rev. Lett. \textbf{96}, 256602 (2006);
	\textbf{98}, 076602 (2007).
\bibitem{AleinerEfetov}
	I. L. Aleiner and K. B. Efetov, Phys. Rev. Lett. \textbf{97}, 236801 (2006);
	J. P. Robinson, H. Schomerus, L. Oroszl\'any, and V. I. Fal'ko, 
	arXiv:0808.2511 [cond-mat.mes-hall] (2008).
\bibitem{OstrovskyGornyiMirlin1}
	P. M. Ostrovsky, I. V. Gornyi, and A. D. Mirlin, Phys. Rev. B \textbf{74}, 235443 (2006).
\bibitem{DasSarma}
	E. H. Hwang, S. Adam, S. Das Sarma, Phys. Rev. Lett. \textbf{98}, 186806 (2007);
	S. Adam, E. H. Hwang, V. M. Galitski, and S. Das Sarma, Proc. Natl. Acad. Sci. U.S.A.
	\textbf{104}, 18392 (2007).
\bibitem{ClassesAII/A}
	P. M. Ostrovsky, I. V. Gornyi, and A. D. Mirlin, Phys. Rev. Lett. \textbf{98}, 256801 (2007); 
	J. H. Bardarson, J. Tworzydlo, P. W. Brouwer, and C. W. J. Beenakker, \textit{ibid.} \textbf{99}, 106801 (2007); 
	S. Ryu, C. Mudry, H. Obuse, and A. Furusaki, \textit{ibid.} \textbf{99}, 116601 (2007);
	K. Nomura, M. Koshino, and S. Ryu, \textit{ibid.} \textbf{99}, 146806 (2007);
	K. Nomura, S. Ryu, M. Koshino, C. Mudry, and A. Furusaki, \textit{ibid.} \textbf{100}, 246806 (2008). 
\bibitem{CheianovFalkoAltshulerAleiner}
	V. V. Cheianov, V. I. Fal'ko, B. L. Altshuler, I. L. Aleiner, Phys. Rev. Lett. \textbf{99}, 176801 (2007).
\bibitem{Kashuba}
	A. Kashuba, arXiv:0802.2216 [cond-mat.mtrl-sci] (2008).
\bibitem{SachdevGrapheneQCC}
	L. Fritz, J. Schmalian, M. M\"uller, and S. Sachdev, Phys. Rev. B \textbf{78}, 085416 (2008). 
\bibitem{SachdevGrapheneHYDRO}
	M. M\"uller, L. Fritz, and S. Sachdev, Phys. Rev. B \textbf{78}, 115406 (2008),
	Markus M\"uller and Subir Sachdev, \textit{ibid.} \textbf{78}, 115419 (2008).
\bibitem{DGKP1}
	M. S. Foster and I. L. Aleiner, Phys. Rev. B \textbf{77}, 195413 (2008).
\bibitem{footnote-a}
	The bulk electrical dc conductivity of clean graphene at the Dirac point $\sigin$ [Eqs.~(\ref{sigmaBulk}) 
	and (\ref{siginDef})] has been calculated in Refs.~\onlinecite{Kashuba,SachdevGrapheneQCC} to the lowest 
	order in the effective electron-electron interaction strength, while large-$N$ work\cite{DGKP1} suggests 
	the possibility of a universal conductivity for moderate to large interaction strengths, appropriate to 
	high temperatures.
\bibitem{AbrikosovBeneslavskii}
	A. A. Abrikosov and S. D. Beneslavskii, Sov. Phys. JETP \textbf{32}, 699 (1971).
\bibitem{StauberGuineaVozmediano}
	T. Stauber, F. Guinea, and M. A. H. Vozmediano, Phys. Rev. B \textbf{71}, 041406(R) (2005).
\bibitem{Son}
	D. T. Son, Phys. Rev. B \textbf{75}, 235423 (2007).
\bibitem{LLv10}
	E. M. Lifshitz and L. P. Pitaevskii, \textit{Physical Kinetics}
	(Pergamon Press, London, 1981).
\bibitem{LofwanderFogelstrom}
	T. L\"ofwander and M. Fogelstr\"om, Phys. Rev. B \textbf{76}, 193401 (2007).
\bibitem{PeresSantosStauber}
	N. M. R. Peres, J. M. B. Lopes dos Santos, and T. Stauber, 
	Phys. Rev. B \textbf{76}, 073412 (2007). 
\bibitem{footnote-b}
	We employ the terms ```generalized' Wiedemann-Franz law'' and ```generalized' Mott
	relation'' to refer to the integral expressions\cite{AshcroftMermin} respectively relating 
	the thermal conductivity $\kappa$ and thermopower $\TEP$ to the bulk dc electrical conductivity $\sigma$, 
	in the ``disorder-limited'' transport regime ($\tauel \lesssim \tauin$).
	In this regime, $\sigma$, $\kappa$, and $\alpha$ can be computed via the Kubo formula
	within the single particle (non-interacting) approximation, although crucial
	renormalization effects must in general be included in the energy-dependence of $\sigma$.\cite{DGKP1}
	In the degenerate limit with $k_{B} \bar{T} \ll |\bar{\mu}|$, the integral expression\cite{AshcroftMermin} 
	relating $\kappa$ ($\alpha$) to $\sigma$ reduces to the algebraic (differential) relation
	conventionally termed the Wiedemann-Franz law (Mott relation). 
	Our primary interest in this paper is the opposite \emph{non}-degenerate limit  
	($k_{B} \bar{T} \gg |\bar{\mu}|$), where the latter
	expressions (and the Sommerfeld expansion) break down.\cite{LofwanderFogelstrom} 
	The ``generalized'' integral relations hold throughout the regime of ``disorder-limited'' 
	transport, and it is to these expressions to which we refer.
\bibitem{AshcroftMermin}
	N. W. Ashcroft and N. D. Mermin, \textit{Solid State Physics}, (Saunders College, Fort Worth, 1976).
\bibitem{Vasko}
	A. Satou, F. T. Vasko, and V. Ryzhii, arXiv:0807.1590 [cond-mat.mtrl-sci] (2008);
	P. N. Romanets, F. T. Vasko, and M. V. Strikha, arXiv:0808.3146 [cond-mat.mtrl-sci] (2008).
\bibitem{Rojo}
	For a review, see e.g.\ A. G. Rojo, J. Phys. Condens. Matter \textbf{11}, R31 (1999).
\bibitem{LLv6}
	L. D. Landau and E. M. Lifshitz, \textit{Fluid Mechanics}
	(Pergamon Press, London, 1959).
\bibitem{footnote-c}
	In our conventions, $g^{i j} \rightarrow \mathrm{diag}(1,-1,-1)$ and $U_{i} U^{i} = v_{F}^2$.
\bibitem{deGroot}
	S. R. de Groot, \textit{Thermodynamics of Irreversible Processes} 
	(North-Holland, Amsterdam, 1963).
\bibitem{Uhlenbeck}
	For a very clear discussion of the systematic extraction of hydrodynamic equations order
	by order in $\tauin$ from the kinetic equation, and the connection to the Chapman-Enskog expansion, 
	see chapters IV and VI in: G. E. Uhlenbeck, G. W. Ford, and E. W. Montroll, 
	\textit{Lectures in Statistical Mechanics}
	(American Mathematical Society, Providence, 1963).
\bibitem{Domenicali}
	C. A. Domenicali, Rev. Mod. Phys. \textbf{26}, 237 (1954).
\bibitem{Konin} 
	A. Konin, Lith. J. Phys. \textbf{46}, 233 (2006).
\bibitem{GrapheneExpt}
	K. S. Novoselov, A. K. Geim, S. V. Morozov, D. Jiang, Y. Zhang, S. V. Dubonos, 
		I. V. Grigorieva, and A. A. Firsov, Science \textbf{306}, 666 (2004);
	K. S. Novoselov, A. K. Geim, S. V. Morozov, D. Jiang, M. I. Katsnelson, I. V. Grigorieva, 
		S. V. Dubonos, and A. A. Firsov, Nature \textbf{438}, 197 (2005);
	Y. Zhang, Y.-W. Tan, H. L. Stormer, and P. Kim, \textit{ibid.} \textbf{438}, 201 (2005);
	Y.-W. Tan, Y. Zhang, H. L. Stormer, and P. Kim, Eur. Phys. J. Spec. Top. \textbf{148}, 15 (2007).
\bibitem{footnote-d}
	The relative chemical potential $\bar{\mu}$ [Eq.~(\ref{muDefs})] is completely determined by
	the relative carrier density $\rho/e = n_{e} - n_{h}$ and the temperature $\bar{T}$. For
	the estimates given in Sec.~\ref{Sec: Numbers}, we have used formulae appropriate to the
	ideal quantum relativistic gas, taking into account valley and spin degeneracies in graphene. 
\bibitem{MounetMarzari}
	See e.g. N. Mounet and N. Marzari, Phys. Rev. B \textbf{71}, 205214 (2005), and
	references therein.
\bibitem{OnoSugihara}
	S. Ono and K. Sugihara, 
	J. Phys. Soc. Jpn. \textbf{21}, 861 (1966);
	K. Sugihara, Phys. Rev. B \textbf{28}, 2157 (1983).
\bibitem{e-phLifetime}
	T. Stauber, N. M. R. Peres, and F. Guinea, Phys. Rev. B \textbf{76}, 205423 (2007);
	F. T. Vasko and V. Ryzhii, \textit{ibid.} \textbf{76}, 233404 (2007);
	E. H. Hwang and S. Das Sarma, \textit{ibid.} \textbf{77}, 115499 (2008).
\end{thebibliography}
\end{document}